\documentclass[preprint,12pt]{elsarticle}
\usepackage{psfig,epsf,rotating,aalongtable,txfonts}

\journal{Icarus}

\begin{document}

\begin{frontmatter}

\baselineskip=25pt
\textwidth 15.0truecm
\textheight 21.0truecm
\topmargin 0.2in
\headsep 1.2cm

\title{Spectroscopic survey of  M--type asteroids\footnote{Based on observations
carried out at the European Southern
Observatory (ESO), La Silla, Chile, ESO proposals 073.C-0622, 074.C-0049 and
078.C-0115 and at the Telescopio Nazionale Galileo, Canary Islands, Spain.}}
\author{S. Fornasier$^{1,2}$, B. E. Clark$^{3,1}$, E. Dotto$^{4}$, \\
 A. Migliorini$^{5}$, M. Ockert-Bell$^{3}$, M.A. Barucci$^{1}$}

\maketitle
\noindent
$^1$ LESIA, Observatoire de Paris, 5 Place Jules Janssen, F-92195 Meudon
Principal Cedex, France\\
$^2$ Universit\'{e} de Paris 7 {\it Denis Diderot}, France \\
$^3$ Department of Physics, Ithaca College, Ithaca, NY 14850, USA \\
$^4$ INAF, Osservatorio Astronomico di Roma, via Frascati 33, I-00040
Monteporzio Catone (Roma), Italy \\
$^5$ IASF-INAF, via del Fosso del Cavaliere 100, 00133 Roma, Italy \\
\noindent
Submitted to Icarus: September 2009\\
e-mail: sonia.fornasier@obspm.fr; fax: +33145077144; phone: +33145077746\\
Manuscript pages: 60; Figures: 10; Tables: 6 \\
\vspace{3cm}

{\bf Running head}: Investigation of M--type asteroids

\noindent

{\it Send correspondence to:}\\
Sonia Fornasier  \\
LESIA-Observatoire de Paris  \\
Batiment 17 \\
5, Place Jules Janssen \\
92195 Meudon Cedex \\
France\\
e-mail: sonia.fornasier@obspm.fr\\
fax: +33145077144\\
phone: +33145077746\\

\newpage
\vspace{2.5cm}

\begin{abstract}

M-type asteroids, as defined in the Tholen taxonomy (Tholen, 1984), are medium
albedo bodies supposed to have a metallic composition and to be the progenitors
both of differentiated iron-nickel meteorites and enstatite chondrites. We carried out 
a spectroscopic survey in the visible and near infrared wavelength range
(0.4-2.5 $\mu$m) of 30 asteroids chosen from the population of asteroids initially classified as Tholen M-types, aiming to investigate their surface composition. The data
were obtained during several observing runs during the years 2004-2007 at the
TNG, NTT, and IRTF telescopes. We computed the spectral slopes in several
wavelength ranges for each observed asteroid, and we searched for diagnostic
spectral features.  We confirm a large variety of spectral behaviors
 for these objects as their spectra are extended into the near-infrared, including the identification of weak absorption bands, mainly of the 0.9
$\mu$m band tentatively attributed to orthopyroxene, and of the 0.43 $\mu$m band that may be associated to chlorites and Mg-rich serpentines or pyroxene minerals such us pigeonite or augite.
A comparison with previously published data indicates that the surfaces of several asteroids belonging to the M-class may vary significantly. \\
We attempt to constrain the asteroid surface compositions of our sample by looking for meteorite spectral analogues in the RELAB database and by modelling with
geographical mixtures of selected meteorites/minerals. We confirm that iron
meteorites, pallasites, and enstatite chondrites are the best matches to most objects 
in our sample, as suggested for M-type asteroids. For 22 Kalliope, we demonstrate that a syntetic mixture obtained enriching a pallasite meteorite with small amounts (1-2\%) of silicates well reproduce the spectral behaviour including the observed 0.9 $\mu$m feature.

The presence of subtle absorption features on several asteroids confirms that
not all objects defined by the Tholen M-class have a pure metallic composition.
 
A statistical analysis of spectral slope distribution versus orbital parameters
shows that our sample originally defined as Tholen M-types tend to be dark in albedo and red in slope for increasing value of the semi-major axis. 
However, we note that our sample is statistically limited by our number of
objects (30) and slightly varying results are found for different subsets. If confirmed, the albedo and slope trends could be due 
to a difference in composition of objects belonging to the
outer main belt, or alternatively to a combination of surface composition, grain
size and space weathering effects.

\end{abstract}

\end{frontmatter}

\newpage
\section{Introduction}

Within the taxonomic system of Tholen (1984), M asteroids have featureless spectra and red spectral slope over the visible wavelength range (0.34 -- 1.04 $\mu$m) and medium visual albedos (0.1 -- 0.3). Tholen and Barucci (1989) identify approximately 40 M-type asteroids using the Tholen (1984) class definition.  The M-types are presumed to be the progenitors both of differentiated
iron-nickel
meteorites and enstatite chondrites (Gaffey 1976; Cloutis et al. 1990; Gaffey et
al. 1993). 
Some authors have proposed alternative meteorite linkages, such as 
CH, CB (Bencubbinite)  (Hardersen et al. 2005), or other carbonaceous chondrites
(Vilas 1994). M-class albedos generally match those of nickel-iron meteorites,
leading some to propose that these asteroids are the collisionally-produced
fragments
of differentiated metal cores (Bell et al. 1989).  
This interpretation requires parent
bodies heated to at least 2000 $^{\circ} C$ to produce iron meteorites (Taylor,
1992).  However, based on recent findings, these interpretations are being
challenged.  Polarimetric measurements (Lupisko \& Belskaya 1989) and radar 
observations (Ostro et al. 1991; 2000; Magri et al. 2007; Shepard et al. 2008) 
of selected M-type asteroids have revealed a large variety of surface
properties that are in some cases inconsistent with only a metallic composition.
Shepard et al. (2008), in a radar survey of 14 M-type asteroids, find that only
four objects (16 Psyche, 216 Kleopatra, 758 Mancunia, and 785 Zwetana) 
have radar albedos high enough to be dominantly metallic. \\
The first estimations of the bulk density of M-asteroids 16 Psyche 
(1.4--1.8  $g/cm^{3}$) and 22 Kalliope (2.4 $g/cm^{3}$) gave values 
surprisingly lower than expected (Viateau 2000; Kocheva 
2003; Britt et al. 2002; Margot \& Brown 2003). These low densities are not 
consistent with predominantly metal composition, and require very high 
values for the bulk porosity. However, these values have been disputed and 
recently Shepard et al. (2008) estimates a very high bulk density 
for 16 Psyche (7.6$\pm$3.0 $g/cm^{3}$), more consistent with a metal
composition. Descamps 
et al. (2008) find a bulk density of 3.35$\pm$0.33 $g/cm^{3}$ for 22 Kalliope,
higher than the first estimation but still implying a high bulk porosity.

Of note is the evidence for hydrated minerals on the 
surfaces of some  M-type asteroids (Jones et al. 1990; Rivkin et al. 1995; 2000;
2002), 
inferred from  the identification of an absorption feature around 3 $\mu$m
typically due 
to the first overtone of H$_{2}$O and to OH vibrational fundamentals in hydrated
silicates. 
Rivkin et al. (1995; 2000) proposed splitting the M-types into {\it wet} W-class
asteroids 
(those which show the 3 $\mu$m band), and {\it dry} M-class asteroids 
(those which do not show the 3 $\mu$m band). 
If the 3 $\mu$m band on M-type asteroids is due to hydrated minerals, 
then these objects might not be anhydrous and igneous as previously believed
and 
the thermal scenarios for the inner main belt may need a revision (e.g. Bell
et al. 1989). 
Gaffey et al. (2002) propose alternative explanations for 
the 3 $\mu$m band: materials normally considered to be anhydrous containing
structural 
OH; troilite, an anhydrous mineral that shows a feature at 3 $\mu$m, 
or xenolithic hydrous meteorite components on asteroid surfaces from impacts 
and solar wind implanted H.

Recent spectroscopic observations in the visible and near infrared region show
that the spectra of M asteroids are not uniformly featureless as previously
believed. In addition to the 3 $\mu$m band, faint
absorption bands in the 0.9 $\mu$m region have been identified on the surfaces
of
some M-type asteroids (Chapman \& Gaffey, 1979; Hardersen et al. 2005; Clark et
al. 2004a; Ockert-Bell
et al. 2008), and attributed to orthopyroxene.  Busarev (1998) detected weak
features tentatively attributed to pyroxenes (0.51 $\mu$m) and oxidized or
aqueously altered mafic silicates (0.62 and 0.7 $\mu$m) in the spectra of the
M-type asteroids 75 Eurydike and 201 Penelope (Busarev 1998).

In this paper we present new spectroscopic data obtained in the visible and
near 
infrared wavelength range (0.4-2.5 $\mu$m) of 30 asteroids sampled from the population of objects whose visible wavelength and albedo properties satisfy the definition of the Tholen M-class. In this work, we (1) identify faint absorption bands which may help reconstruct surface composition, (2) search for meteorite spectral analogues, and (3)
model surface composition with geographical mixtures of selected
minerals when a meteorite match is not found.

\section{Data Acquisition}  

[HERE TABLE 1]

The data presented in this work were obtained  during 2 runs 
(February and November 2004) at the italian Telescopio Nazionale Galileo
(TNG) of the European Northern Observatory (ENO) in la Palma, Spain, and 3 runs
(May 2004, August 2005, and January 2007) at the New Technology Telescope (NTT)
of the European Southern Observatory (ESO), in Chile. Six asteroids,
investigated in the visible range at the NTT and TNG telescopes, were separately
observed in the near infrared at the Mauna Kea Observatory  3.0 m NASA Infrared
Telescope Facility (IRTF) in Hawaii during several 
runs in 2004-2008 (Table ~\ref{tab1}).
Our survey was devoted to the investigation of asteroids 
belonging to the E, M and X classes, following the Tholen (1984) taxonomy. 
Here we present the results of our observations sampling within the Tholen M-class. The results from Tholen E-type asteroids were published in Fornasier et al. (2008) and the results from Tholen X-type observations will be presented in a separate paper.

\subsection{Observations at the TNG Telescope}
At the TNG telescope, for visible spectroscopy we used the Dolores (Device
Optimized for the 
LOw RESolution) instrument equipped with the low resolution red grism 
(LR-R) covering the 0.51--0.95 $\mu$m range with a spectral dispersion 
of 2.9 \AA/px (see $http://www.tng.iac.es/instruments/lrs$). Most of the objects
were also observed with the low resolution blue grism 
(LR-B, dispersion of 2.8 \AA/px) in February 2004, covering the 0.4-0.8 $\mu$m
range, and with the medium resolution blue grism (MR-B, dispersion of 1.7
\AA/px, 0.4-0.7 $\mu$m range), for the November 2004 run. The 'red' and 'blue'
spectra in the visible range were separately reduced and finally combined
together to obtain spectral coverage from 0.40 to 0.95 $\mu$m.
The Dolores detector is a Loral thinned and 
back-illuminated CCD with 2048 $\times$ 2048 pixels, with a pixel size of 
15 $\mu$m and a pixel scale of 0.275 arcsec/px. Like most of the Loral CCDs, 
the Dolores chip is affected by moderate-to-strong fringing at red wavelengths. 
Despite taking as much care as possible in the data reduction process, some 
asteroid spectra show residual fringing that impedes identification 
of absorption bands in the 0.9 $\mu$m region.

For the infrared spectroscopic investigation at the TNG telescope we used the
near 
infrared camera and spectrometer (NICS) equipped with an Amici prism disperser. 
This equipment covers the 0.85--2.40 $\mu$m range during a single 
exposure with a spectral resolution of about 35 (Baffa et al. 2001). 
The detector is a 1024 $\times$ 1024 pixel Rockwell HgCdTe Hawaii array. 
The spectra were acquired nodding the object
along the spatial direction of the slit, in order to obtain alternating pairs 
(named A and B) of near--simultaneous images for the  background removal. 
For both the visible and near infrared observations we utilized a 1.5 arcsec 
wide slit, oriented along the parallactic angle to minimize 
the effect of atmospheric differential refraction.

\subsection{Observations at the NTT Telescope}
At the NTT telescope, visible spectra were acquired using the EMMI instrument, 
equipped with a 2x1 mosaic of 2048$\times$4096 MIT/LL CCD with square 15$\mu$m
pixels (D'odorico, 1990).
We used the grism \#1 (150 gr/mm) in RILD mode to cover the wavelength 
range 4100--9600 \AA\ with a dispersion of 3.1 \AA/px at the first order. 
For the near infrared spectroscopy, we utilized the instrument SOFI (Son OF
Isaac) 
in the low resolution mode. The  blue and red grisms, covering respectively the 
0.95--1.64 $\mu$m (dispersion of 6.96 \AA/px) and the 1.53--2.52 $\mu$m range 
(dispersion of 10.22 \AA/px) were used (Moorwood et al., 1998). The
acquisition technique was identical to 
that used for the TNG observations, however we used a larger slit (2 arcsec), 
always oriented along the parallactic angle.
The two spectral regions of the blue and red grism observations were combined
matching 
the overlapping region in the 1.53--1.64 $\mu$m range.  
%
\subsection{Observations at the IRTF Telescope}
At the IRTF telescope we used the SpeX instrument, equipped with a cooled
grating and an InSb array (1024 x 1024) spectrograph at wavelengths from 0.82 to
2.49 $\mu$m. (Rayner et al. 2003).  Spectra were recorded with a slit oriented
in the north-south direction and opened to 0.8 by 15 arcseconds. A dichroic lens
reducing the signal below 0.8 $\mu$m was used for all observations.  Objects
were consistently observed near the meridian to minimize airmass, but total
integration time varied from $\sim$ 6 to 30 minutes, depending on the
strength of the signal relative to sky lines.

\section{Data Reduction}
Visible and near-infrared spectra were reduced using ordinary procedures 
of data reduction with the software packages Midas and IDL as described in
Fornasier et al. (2004a,b). 
For the visible spectra, the procedure includes the subtraction of the bias 
from the raw data, flat--field correction, cosmic ray removal, sky subtraction,
collapsing the
two--dimensional spectra to one dimension, wavelength calibration,
and atmospheric extinction correction. The reflectivity of each asteroid was 
obtained by dividing its spectrum by that of the solar analog star closest in
time and airmass to
the object. Spectra were finally smoothed with a median filter 
technique, using a box of 19 pixels in the spectral direction for each point of
the 
spectrum. The threshold was set to 0.1, meaning that the original value was
replaced by 
the median value if the median value differs by more than 10\% from the original
one.

For observations in the infrared range, spectra were first corrected for flat
fielding, 
then bias and sky subtraction
was performed by producing A-B and B-A frames. The positive spectrum of the B-A
frame was 
shifted and added to the positive spectrum of 
the A-B frame. The final spectrum is the result of the mean of all pairs of 
frames previously combined. The spectrum was extracted and wavelength 
calibrated. For the NTT-SOFI spectra, this last step was performed acquiring 
the spectrum in the blue and red grisms of a Xenon lamp, and comparing the
observed 
lines with those of a reference table, obtaining the dispersion relation.
For the TNG-NICS spectra, due to the very low resolution of the Amici prism, 
the lines of Ar/Xe lamps are blended and 
cannot be easily used for standard reduction procedures. 
For this reason, the wavelength calibration was obtained using a look-up table
which is 
based on the theoretical dispersion predicted by ray-tracing and adjusted 
to best fit the observed spectra of calibration sources.
Finally, the extinction correction and solar removal was obtained by 
division of each asteroid spectrum by that of the solar analog star closest in
time and airmass to
the object. 
For the IRTF data, after the normal data reduction procedures of 
flat-fielding, sky subtraction, spectrum extraction, and wavelength 
calibration, each spectrum was fitted with the ATRAN atmospheric 
model for telluric absorption features (Lord 1992; Bus et al. 2002a, 2003; 
Sunshine et al. 2004).  This procedure required an initial estimate 
of precipitable water in the atmospheric optical path using the zenith 
angle for the observation and the known $\tau-$values (average atmospheric 
water) for Mauna Kea.  This initial guess was iterated until the best 
fit between predicted and observed telluric band shapes was obtained, 
and an atmospheric model spectrum was generated (Bus et al. 2003).  
Following this, each asteroid spectrum was divided by the atmospheric 
model and then ratioed to each star spectrum, similarly reduced, before 
normalization at 1.2 $\mu$m.  The software and procedures used for data 
reduction are described in DeMeo et al. (2009).  The final spectra we 
report are averages of 3-5 asteroid/star ratios, calculated to minimize 
variations due to standard star and sky variability.  Usually, 2-5 
different standard stars were observed on any given night at the telescope. 
We used only “solar” standard stars, and each standard star was observed 1-3
times per night.

The stellar and asteroid spectra were cross-correlated and, if necessary, 
sub-pixel shifts were made before ratioing the asteroid to the star. 
This reduction step is needed to reduce the noise and/or the changes in the 
final asteroid slope due to small changes in the wavelength dispersion between 
asteroid and star observations, introduced by instrumental flexure. 

The infrared and visible spectral ranges of each asteroid were finally combined
by overlapping the spectra, merging the two wavelength regions at the common
wavelengths and utilizing the zone of good atmospheric transmission to find the
normalization factor between the two spectral parts. For the TNG and IRTF data
the overlapping region goes from $\sim$ 0.86 to 0.94 $\mu$m, and we took the
average value over the 0.89-0.91 $\mu$m region of the visible spectrum to
normalize the infrared spectrum. For the NTT data, the overlapping region is
very
small so we took the average value over the 0.95-0.96 $\mu$m region to
normalize the spectra. 
The spectra of the observed asteroids, all normalized at 0.55 $\mu$m, are shown
in Figures \ref{fig1}--\ref{fig5}.
Observational conditions are reported in Table~\ref{tab1}.

[Here Figs 1, 2, 3, 4, 5]

\section{Results: Spectral Analysis and Absorption Features}

[HERE TABLE 2 and 3]

\begin{figure*}
\centerline{\psfig{file=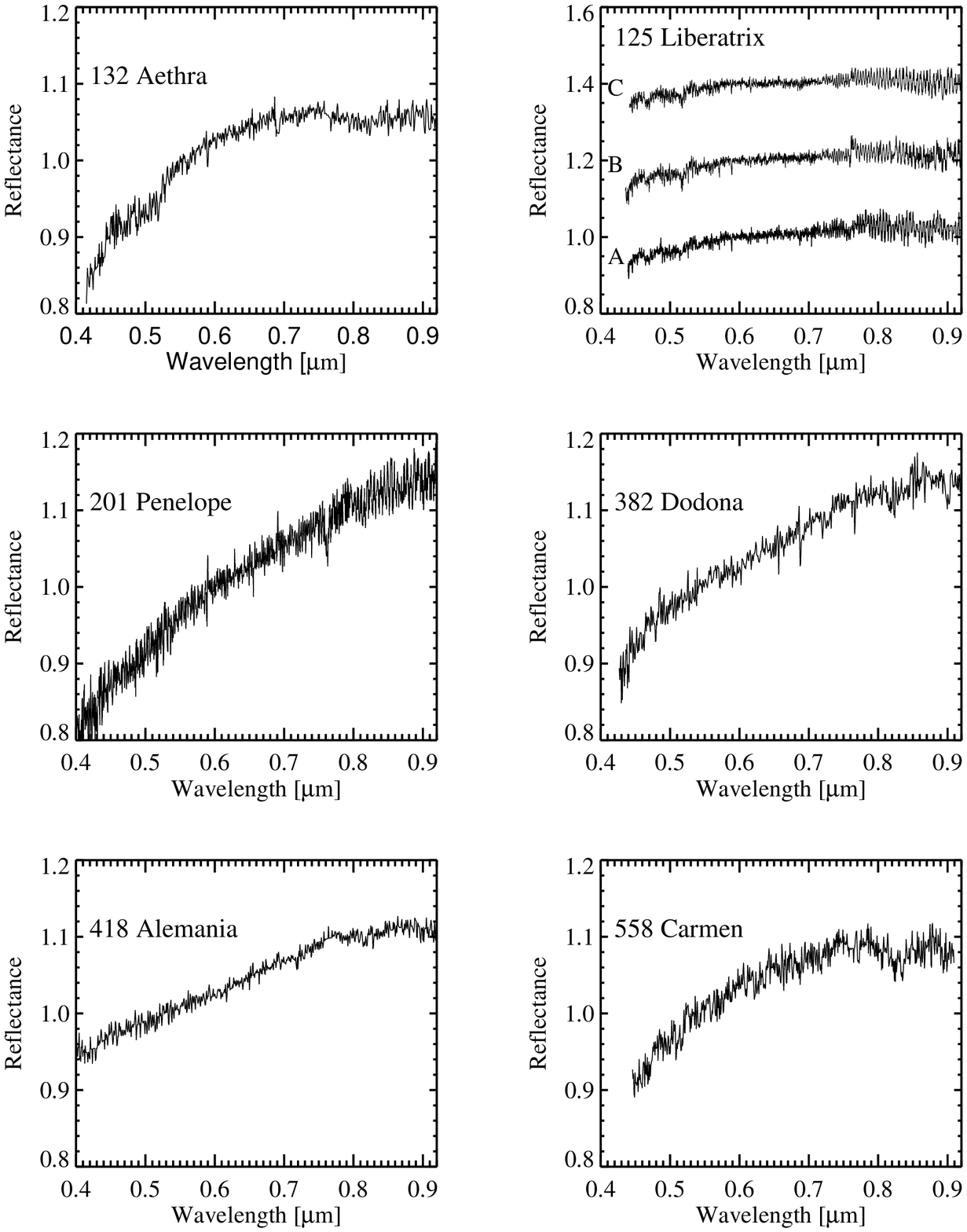,angle=0,width=16.5cm}}
\caption{}
\label{fig1}
\end{figure*}

\begin{figure*}
\centerline{\psfig{file=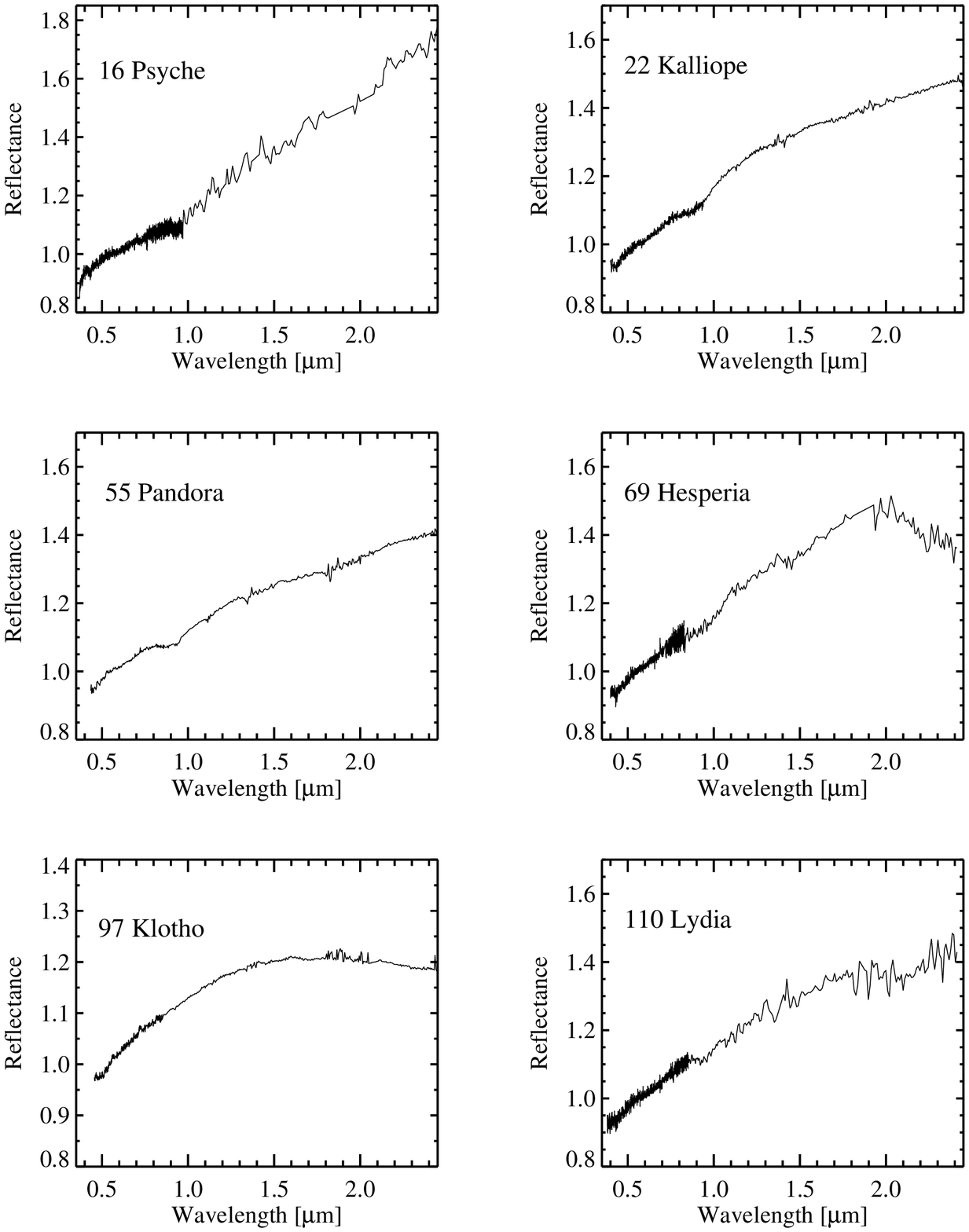,angle=0,width=18.5cm}}
\caption{}
\label{fig2}
\end{figure*}

\begin{figure*}
\centerline{\psfig{file=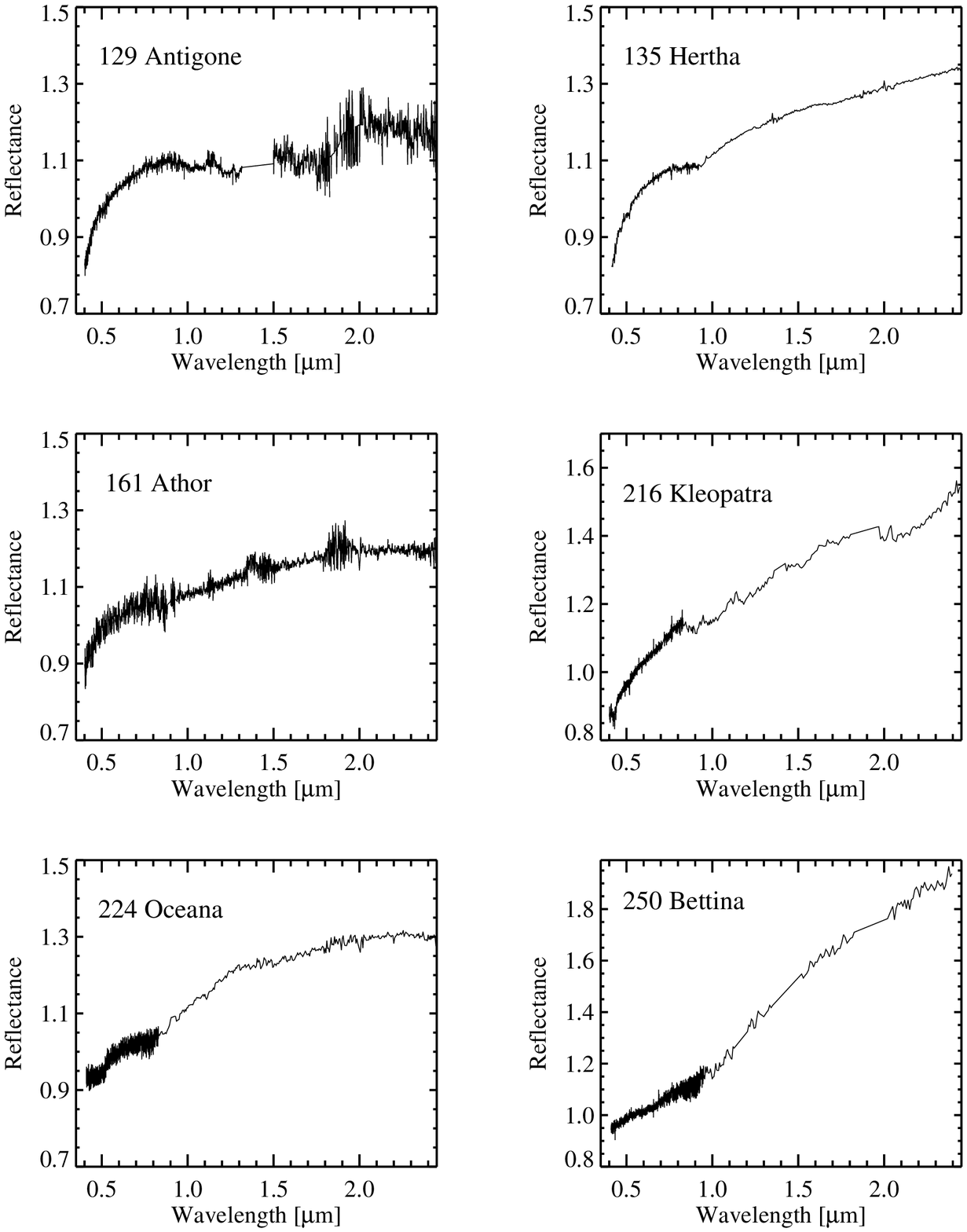,angle=0,width=18.5cm}}
\caption{}
\label{fig3}
\end{figure*}

\begin{figure*}
\centerline{\psfig{file=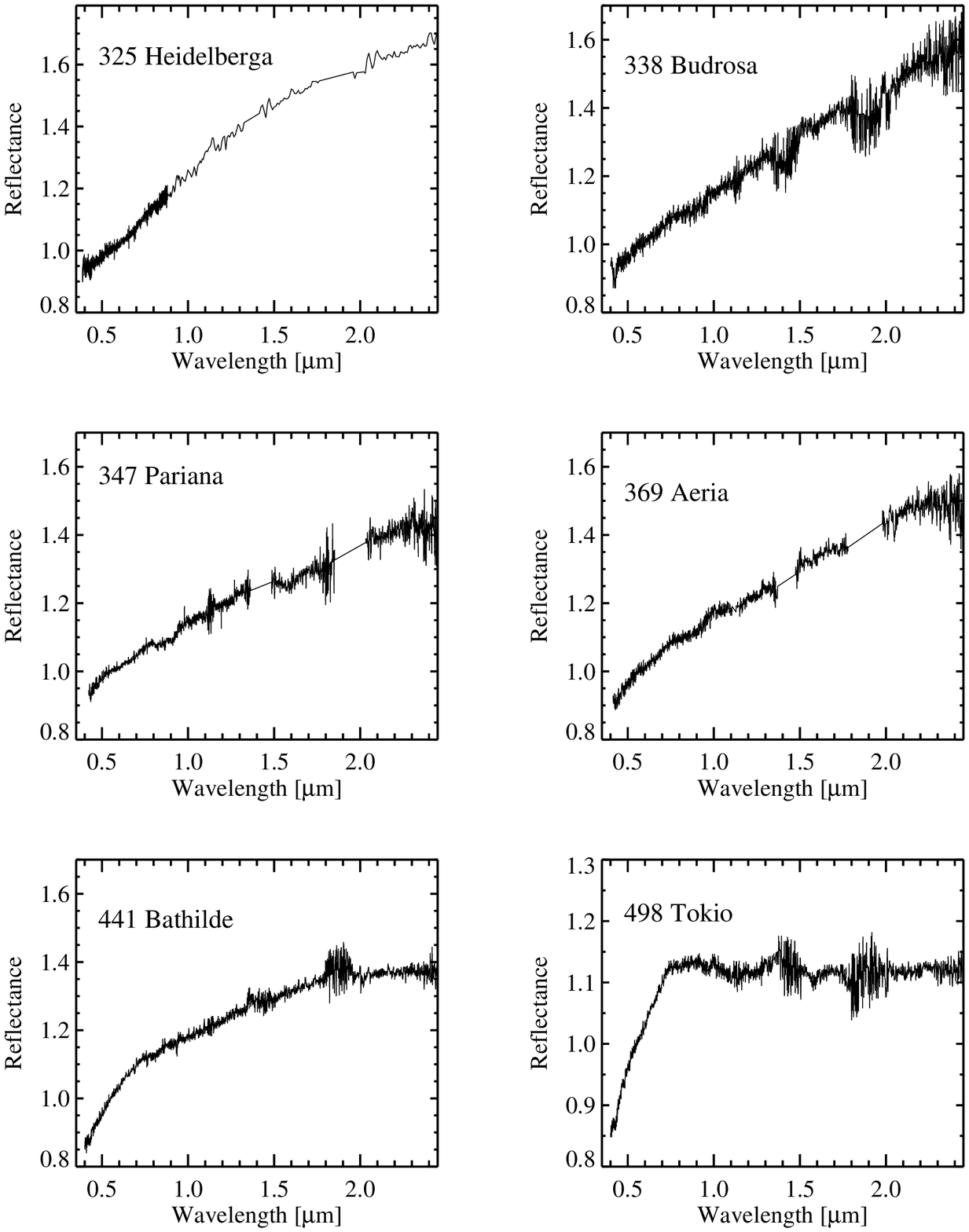,angle=0,width=18.5cm}}
\caption{}
\label{fig4}
\end{figure*}

\begin{figure*}
\centerline{\psfig{file=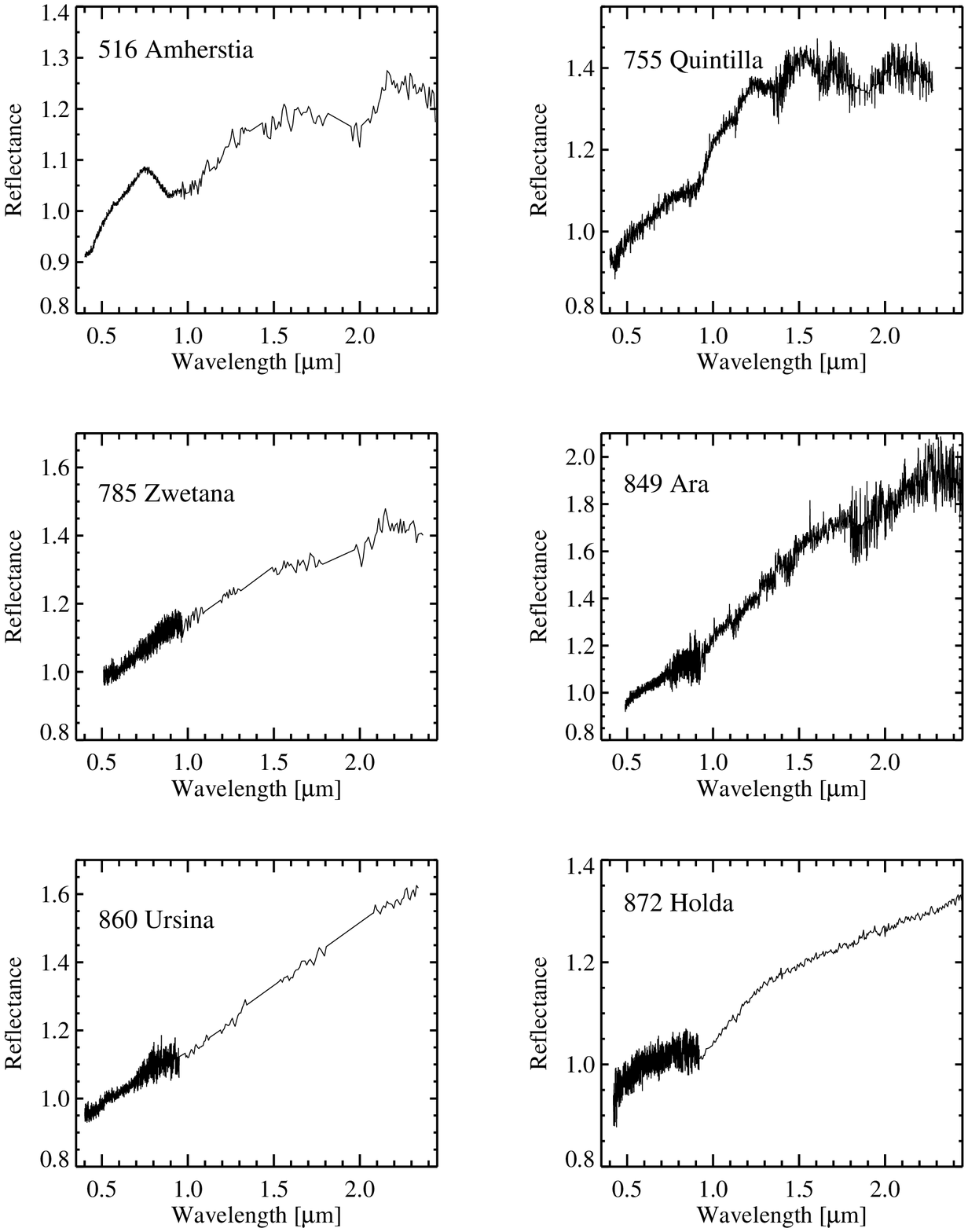,angle=0,width=18.5cm}}
\caption{}
\label{fig5}
\end{figure*}

We present new spectra of 30 asteroids whose visible wavelength properties and albedos fall within the boundaries of the M-class as defined by Tholen (1984).
We obtained new visible spectra for 6 objects, 
and new visible and near-infrared spectra for 24 objects 
(Figures \ref{fig1}--\ref{fig5}).

To analyze the data, spectral slope values were calculated by a linear
fit to different wavelength regions: $S_{cont}$ is the spectral slope in the
whole
range (continuum slope) observed for each asteroid, S$_{VIS}$ is the slope in
the 0.55-0.80 $\mu$m
range, S$_{NIR1}$ is the slope in the 1.1-1.6 $\mu$m range, and S$_{NIR2}$ is
the slope in the 1.7-2.4 $\mu$m range. Values are reported in
Table~\ref{slope}, together with the asteroid's physical and orbital parameters,
and the Bus-DeMeo classifications (Bus \&Binzel 2002b; DeMeo et al. 2009)
derived respectively from the visible and VIS-NIR spectrum. 
Band centers and depths were calculated for each asteroid showing an 
absorption feature, following the Gaffey et al. (1993) method. 
First, a linear continuum was fitted at the edges of the band, 
that is at the points on the spectrum outside the absorption 
band being characterized. Then the asteroid spectrum was divided 
by the linear continuum and the region of the band was fitted with a 
polynomial of order 2 or more. The band center was then calculated 
as the position where the minimum of the spectral reflectance 
curve occurs, and the band depth as the minimum in the ratio of the 
spectral reflectance curve to the fitted continuum (see Table~\ref{band}).

The infrared data reveal the silicatic nature of 516 Amherstia. Its
spectrum exhibits the $\sim$ 1 and 2 $\mu$m bands 
typically associated with olivine and pyroxene (Table~\ref{band}). 
On this basis, we re-classify 516 as an S--type (it belongs to 
the Sq class in the Bus-DeMeo classification).  For this reason, 
516 will not be included in the following discussion.  
The asteroids in our sample  show different spectral behaviors
(Figs~\ref{fig1}--\ref{fig5}). Sixteen of them have spectra with at least one
absorption feature (Table~\ref{band}). The main feature identified is in the 0.9
$\mu$m region, as already reported by several authors 
(Clark et al. 2004a; Hardersen et al. 2005; 2006a,b; 2007a,b;
Ockert-Bell et al. 2008), and it is attributed to low-Fe, low-Ca
orthopyroxene minerals.
The band center ranges from 0.86 $\mu$m (417 Suevia) to 
0.97 $\mu$m (216 Kleopatra), with a band depth of 1--5 \% as compared to the
continuum. 

Some asteroids (16 Psyche, 22 Kalliope, 69 Hesperia, 216 Kleopatra, 338 Budrosa,
and 498 Tokio) show an additional faint feature at $\sim$0.43 $\mu$m
(Table~\ref{band}). This band is not easily attributable: it might be associated
with chlorites and Mg-rich serpentines as suggested by King \& Clark (1989) for 
enstatite chondrites; with pyroxene minerals such us pigeonite or augite as
suggested by Busarev (1998) for M-asteroids; or to an Fe$^{3+}$  spin-forbidden
transition in the iron sulfate jarosite, 
as suggested by  Vilas et al. (1993) for
low-albedo asteroids. Within our sample, Rivkin et al. (1995; 2000) detected the 3 $\mu$m absorption band on 22 Kalliope,
55 Pandora, 110 Lydia, 129 Antigone, 135 Hertha, 201 Penelope (they proposed the
W {\it wet} class for these bodies), and they suggested that this feature can be
attributed to the presence of hydrated phases on their surfaces (Table
~\ref{band}). We detect the 0.43 $\mu$m band only on one W asteroid (22
Kalliope). If this feature is actually due to the presence of chlorites-- Mg
rich
serpentines and/or to jarosite, this might support the interpretation of the 3
$\mu$m band as due to aqueous alteration products. The other W
asteroids do not show the 0.43 $\mu$m feature, while it {\it is} present in
the spectra of two asteroids not having the 3 $\mu$m absorption (16 Psyche and
216 Kleopatra). We conclude that there is no correlation between the 
0.43 and 3 $\mu$m bands. 

Asteroid 132 Aethra shows a faint band centered at $\sim$ 0.49 $\mu$m that
resembles the absorption seen on some E-type (subclass EII) asteroids (Fornasier
et al. 2007; 2008; Clark et al. 2004b), where it is attributed to sulfides such
as oldhamite and/or troilite (Burbine et al. 1998; 2002a; 2002b). We obtained
only one spectrum in the visible range and 
no other observations are available in
the near infrared nor in the 3 $\mu$m region. Bus and Binzel (2002b) 
gave 132 Aethra an Xe designation, and its IRAS albedo is 0.2 (Table 4).  

Asteroid 135 Hertha has a faint band at $\sim$ 0.51 $\mu$m, narrower than
that seen on EII asteroids or Aethra, and similar to the Fe$^{2+}$
spin-forbidden crystal field transitions in Earth and lunar pyroxene (Burns et
al. 1973, Adams 1975, Hazen et al. 1978). This band has been previously detected
by Busarev (1998) on the surface of two other M asteroids, 75 Eurydike and 201
Penelope (however our visible spectrum of 201 does not show any 
feature detectable within the noise of the data).

125 Liberatrix was observed 3 times on November 20, 2004. The three spectra are
shown in Fig.~\ref{fig1}, and were obtained $\sim$ 1.5 hours apart, covering
$\sim$ 75\% of the Liberatrix rotational phase (period of 3.97 hours). The
spectra
are very similar (Fig.~\ref{fig1}) so, in this region of the rotational phase,
the body appears to have a homogeneous surface composition. There is a hint of
an absorption band in the 0.9 $\mu$m region, also seen by Hardersen et
al. (2005), however we cannot clearly identify it due to the aforementioned
fringing in this region. The features at
0.48 and 0.52 $\mu$m  are spurious residuals of the flat field. 

\subsection{Comparison with Existing Literature Data} 

Hardersen et al. (2005) report the presence of absorption bands in the 0.9
$\mu$m
region on 16 Psyche, 69 Hesperia, 110 Lydia and 216 Kleopatra (and an absence of
absorption
features on 325 Heidelberga, within the scatter of their data). Our spectra
confirm
the presence of an absorption feature in the 0.9 $\mu$m region, but our band
center values are higher than those reported by Hardersen et al. (2005). They
found the band also on 125 Liberatrix and 201 Penelope, which we observed only
in the
visible region with the TNG telescope, and which show an indication of a faint
absorption feature (Fig~\ref{fig1}). Nevertheless, due to the fringing problems
and low sensitivity of the detector in this wavelength region, we cannot
clearly confirm that an absorption band is present, nor can we quantify
it. The difference in the band center location might be due to different
methodologies in data analysis (choice of continuum removal or imperfect removal
of the telluric water vapor feature), but we cannot exclude that it is real and
associated with variations in the orthopyroxene content over the asteroid's
surface. 

Birlan et al. (2007) observed 9 M-type asteroids, and they did not report the
presence of absorption features. Three of their objects belong also to our
survey. Our
spectra of 325 Heidelberga and 860 Ursina seem in agreement with the data of
Birlan et al. (2007), while for 558 Carmen we obtained only a spectrum in the
visible range. 558 Carmen was observed
also by Hardersen et al. (2006a) through about 1/3$^{rd}$ of a full rotation,
and their spectra exhibit a weak 0.9 $\mu$m band, attributed to low-Fe
pyroxenes, with band centers ranging from 0.91 to 0.94 $\mu$m. The Birlan et al.
(2007) spectrum is featureless, and our visible spectrum ends at 0.91 $\mu$m so
it can not help with band identification (the faint absorption around
0.83 $\mu$m is due to incomplete removal of telluric bands). It is possible
that asteroid 558 Carmen has an inhomogeneous surface composition. 

325 Heidelberga was observed also by Takir et al. (2008) who found 
in its spectrum a weak (1-2\% depth) feature center at 0.9 $\mu$m, 
while our spectrum, that of Birlan et al. (2007), and of Hardersen et al.
 (2005) are featureless. This asteroid may have a heterogeneous surface 
composition and further high S/N ratio observations covering the whole
rotational 
period are needed for a full investigation. 
 
Hardersen et al. (2006a) detected the $\sim$ 0.9 $\mu$m absorption 
feature also on 216 Kleopatra (center at 0.91 $\mu$m, depth 3\%), and on 
347 Pariana (center varying from 0.88--0.93 $\mu$m with depth  3\%, 
in spectra obtained through more than one asteroid full rotation). 
We confirm the detection of the band on these two asteroids (Table ~\ref{band}) 
with similar band parameters. Our data agree with the Hardersen et al. (2006a) 
finding also for 97 Klotho (featureless spectra in both case, and also for the
Klotho 
data presented in Ockert-Bell et al. 2008), but disagree for 498 Tokio and 129
Antigone. 
In fact, for 498 Tokio, Hardersen et al. (2006a) reported a blue featureless
spectrum, 
while our data show an increasing slope up to $\sim$ 0.7 $\mu$m, and then a flat
spectral 
behavior in the near-infrared, with two faint absorption bands centered at
0.43, and 1.16 $\mu$m. On the other hand, for 129 Antigone Hardersen et al.
(2006a) 
found absorption features at 0.76, 0.9, 1.07, and 1.39 $\mu$m, attributed to the
presence of antigorite, which has absorption bands consistent with those of the
asteroid. 
Our spectrum has a poor S/N ratio in the near-infrared range.
We do not see the 0.76 or 0.9 $\mu$m bands, but a broad feature in the
0.93--1.14 $\mu$m region. 
The spectrum is peculiar: it has a convex shape below 0.9 $\mu$m, then it is
almost 
flat up to $\sim$ 1.8 $\mu$m with a steep increase of the spectral slope 
in the 1.8--2.0 $\mu$m region. Ockert-Bell et al. (2008) observed Antigone at 
different aspects, and they found the 0.9 $\mu$m band and spectral variations 
in continuum slope above 20\% for different rotational aspects. It is thus 
possible that this asteroid has an inhomogeneous surface composition.
  
Hardersen et al. (2007a; 2007b; 2006b) observed several asteroids reported in
this survey (135 Hertha, 224 Oceana, 250 Bettina, 369 Aeria, 516 Amherstia, and
872 Holda). For 250 Bettina, our spectrum is steeper in the infrared compared to
their data, but both spectra show a band in the 0.9 $\mu$m region. For 369
Aeria, Hardersen et al. found absorption bands at 0.9 and 1.9 $\mu$m, while our
spectrum shows only the 0.9 $\mu$m feature and a reddish spectral behavior in
the
NIR with a steeper spectral slope than Hardersen et al. (2007a). 224 Oceana
shows similar featureless spectra and 135 Hertha has an absorption at 0.9 $\mu$m
in both surveys. 516 Amherstia shows similar $\sim$ 1 and 2 $\mu$m features, but
they are deeper in our spectrum compared to Hardersen et al (2006b). 
872 Holda shows a
broad and weak feature in both surveys centered at $\sim$ 0.96 $\mu$m. Hardersen
et al. (2006b) found that Holda is spectrally similar to synthetic troilite
(Cloutis and Burbine 1999), and proposed for the surface composition of this
asteroid a mixture of NiFe metal and troilite. 

Four M-types (16 Psyche, 110 Lydia, 216 Kleopatra, and 785 Zwetana) of the Clark
et al. (2004a) 
survey of the X complex were re-studied in this work. The spectral behavior of 
110 Lydia and 216 Kleopatra looks quite similar, while for 16 Psyche our data
show a 
linear NIR spectrum with an higher spectral gradient than that reported in 
Clark et al. (2004a), and for 785 Zwetana our NIR spectrum is almost linear and 
featureless, while Clark et al. (2004a) reported a convex shape spectrum and 
detected a feature at $\sim 0.9 \mu$m. Psyche was observed also by Binzel et al. (1995) over its rotation, and they did not find any significant variation in its visible reflectance spectrum within their measurement precision of 1\%.
 
 Zwetana was observed also by Hardersen 
et al. (2007a), who detected in its spectrum a broad and faint absorption
beyond 
1.3 $\mu$m, but no features in the 0.9-1.0 $\mu$m region, and by Ockert-Bell 
et al. (2008), who presented a featureless spectrum with no significant slope 
variations at different aspects.  

755 Quintilla shows a very peculiar spectrum with several absorption bands 
(Fig~\ref{fig5},  Table~\ref{band}). The bands at 0.9 and 1.86 $\mu$m can be 
attributed to orthopyroxene, however the peculiar features at 1.37 and 1.61
$\mu$m 
are not easily interpreted. It is worth noting that Fieber-Beyer 
et al. (2006) find three absorption features in a Quintilla spectrum but 
at different central wavelengths. They interpreted bands at 0.93, and 
1.7 $\mu$m as due to spinel, while the nature of the 1.59 $\mu$m band was
unknown. 

In sum, comparing our spectra with those in the existing literature, we 
suggest that asteroids 129 Antigone, 325 Heidelberga, 498 Tokio, and  785
Zwetana may display surface variability as they show different spectral
behaviors throughout independent observations. Some differences in the spectral
slope values or band center positions are seen also for asteroids 16 Psyche, 250
Bettina, 369 Aeria, 516 Amherstia, 558 Carmen, and 755 Quintilla, but we cannot
clearly state if they are due to real variation or simply to observational
uncertainty and/or differences in data acquisition and reduction.  Additional 
high S/N ratio observations are needed to investigate possible surface
inhomogeneities for these asteroids.

\section{From the Tholen to the Bus-DeMeo taxonomy}

[HERE TABLE 4]

Our targets were classified as belonging to the Tholen M class on the basis of
their spectra from 0.4 to 1 $\mu$m and their moderate albedo values. Our
new spectral observations enrich the available physical information for these
objects and allow us to apply the  Bus-DeMeo classification recently published
(DeMeo et al., 2009). The Bus-DeMeo system is based on the 
asteroid's spectral
characteristics over the wavelength range 0.45 to 2.45 $\mu$m 
without taking the albedo into
consideration. Indeed, the observed Tholen M-type asteroids
show different spectral behaviors in the near infrared range. We therefore
re-classified our 24 targets observed in the V+NIR range according to the 
Bus-DeMeo taxonomy and the corresponding classes are summarized in Table
~\ref{slope}. 

Thirteen of the 24 Tholen M-type asteroids we investigated are Xk-types in
the Bus-DeMeo taxonomy. Asteroids belonging to this class have spectra
characterized  by a shallow absorption feature that appears as a concavity over
the range 0.8 to 1.0 $\mu$m, in what is otherwise a generally linear spectrum.
Indeed, all 13 asteroids classified as Xk (Table~\ref{slope}) show
a band in the 0.9 $\mu$m region (Table~\ref{band}) attributed to low-Fe, low-Ca
orthopyroxene minerals. Their albedo value varies between 0.12 and 0.21, except
for 2 asteroids, where it is higher (250 Bettina has an albedo of 0.25 and 55
Pandora of 0.30). Some of the Xk asteroids investigated show additional 
features as summarized in Table~\ref{features}. A 0.43 $\mu$m band is seen in the Xk asteroids 16, 22, 69, 216, and 338. Considering the asteroids' 
moderate albedo values, this band might be due to  chlorites 
and Mg-rich serpentines or pigeonite or augite minerals. \\
For 755 Quintilla, the classification as Xk type seems to be questionable
since Quintilla presents other features at 1.37,
1.61, and 1.87 $\mu$m, as previously mentioned (Table~\ref{features}).
Finally, 135 Hertha also shows a faint feature at 0.51 $\mu$m 
probably due to Fe$^{2+}$
spin-forbidden crystal field transitions in pyroxene. 

Five of the observed asteroids (Table~\ref{slope}) have a slightly reddish
spectrum, featureless in the 0.45-2.4 $\mu$m range, and belong to the Xc-type in
the Bus-DeMeo taxonomy. Among them only 498 Tokio has a low albedo value
compatible with carbonaceous chondrite analogs. It is the only Xc type 
that also show a feature at 0.43 $\mu$m, that we interpret as due to 
the iron sulfate jarosite, and a broad absorption in the 
1-1.3 $\mu$m region usually seen in C-type 
asteroids (Table~\ref{features}). The Xc type 129 
Antigone has a band centered at 1.03 $\mu$m possibly due to the 
phyllosilicate antigorite, as suggested by Hardersen et al. (2006a).  

The four asteroids 325, 441, 785, and 860 have featureless spectra and follow
the X class. Their albedo values vary between 0.11-0.16.

According to these new near infrared observations 516 Amherstia must be
reclassified as an S-type in the Tholen classification and an
Sq-type in the Bus-DeMeo system. 
The asteroid 849 Ara has a steep spectral slope and falls in the D class
according to the Bus-DeMeo taxonomy (Table~\ref{slope}), but its albedo is very high (0.27), excluding a surface composition of organic-rich silicates, 
carbon, and anhydrous silicates as commonly expected on low-albedo 
Tholen D type asteroids.

\section{Meteorite Spectral Matches}

\subsection{Methodology}

To constrain the possible mineralogies of our asteroids (those with both visible
and near 
infrared spectra), we conducted  a search for meteorite and/or 
mineral spectral analogs.  We used the  
publicly available RELAB spectrum library (Pieters 1983), which  
consisted of nearly 15,000 spectra in November of 2008.  For each  
spectrum in the library, a filter was applied to find relevant  
wavelengths (0.4 to 2.45 $\mu$m).  Then, a second filter was applied  
to reject spectra with irrelevant albedo values (i.e. brightness at 0.55  
$\mu$m too high or too low).   
This process produced a list of approximately 4000 RELAB  
spectra of appropriate brightness and wavelength coverage for  
comparison to the asteroid.   RELAB spectra were normalized to 1.0 at  
0.55 $\mu$m, and then a Chi-squared value was calculated relative to  
the normalized input asteroid spectrum. 
We used the asteroid wavelength sampling to resample the laboratory
spectra by linear interpolation.  This allowed a least-squares
calculation with the number of points equal to the wavelength
sampling of the asteroid spectrum.  
The RELAB data files were sorted according  
to Chi-squared, and then visually examined for dynamic weighting of  
spectral features by the spectroscopist.  Given similar Chi-squared  
values, a match that mimicked spectral features was preferred over a  
match that did not.  We visually examined the top 50 Chi-squared  
matches for each asteroid. No specific grain-size  
sampling filter was applied, and many of the best matches found were of
slab or sanded slab (or rough-cut) surfaces.

The importance of our second filter is that we constrained the 
search for analogs to those laboratory spectra with brightness  
(reflectance at 0.55 $\mu$m) roughly comparable to the albedo of the  
asteroid.  For the asteroid albedo value, we generally used the IRAS  
albedo published by Tedesco et al. (2002).  For the darker asteroids  
(below 10\% albedo), this is a fairly well-known number.  The  
uncertainty for brighter asteroid albedos (above 10\% albedo) is  
relatively larger.  This is because at equivalent geocentric and  
heliocentric distances, a darker asteroid will have stronger thermal  
flux, allowing a more precise and more accurate albedo determination.   
To account for this, we filtered out lab spectra if brightness  
differed by more than $\pm$ 3\% in absolute albedo for the darker  
asteroids in our sample (albedo less than 10\%), and we filtered out lab spectra
if  
brightness differed by more than $\pm$ 5\% in absolute albedo for the  
brighter asteroids  (albedos greater than 10\%).  For example, if an  
asteroid’s albedo was 20\%, we searched over all lab spectra with  
0.55 $\mu$m reflectance between 15\% and 25\%.  If an asteroid's albedo
was 7\%, we searched over all lab spectra with reflectance between
4\% and 10\%.  Once the brightness filter was  
applied, all materials were normalized before comparison by least-squares.
We note that mismatches between asteroid albedo and 
meteorite 0.55 $\mu$m reflectance occurs 
for a few well-known meteorite-asteroid pairings, such as Vesta 
(albedo about 40\%) and HEDs (many of which have 0.55 micron 
reflectances of 25-30\%). 
The last filter to be applied was a "no lunar sample" filter,
and this will be explained further in the next section.
Our search techniques effectively emphasized the spectral characteristics of
brightness and shape, and de-emphasized minor absorption bands and other
parametric characteristics.

\subsection{Meteorite Analogs}

[HERE TABLE 5]

[HERE FIGURES 6, 7 and 8]

\begin{figure*}
\centerline{\psfig{file=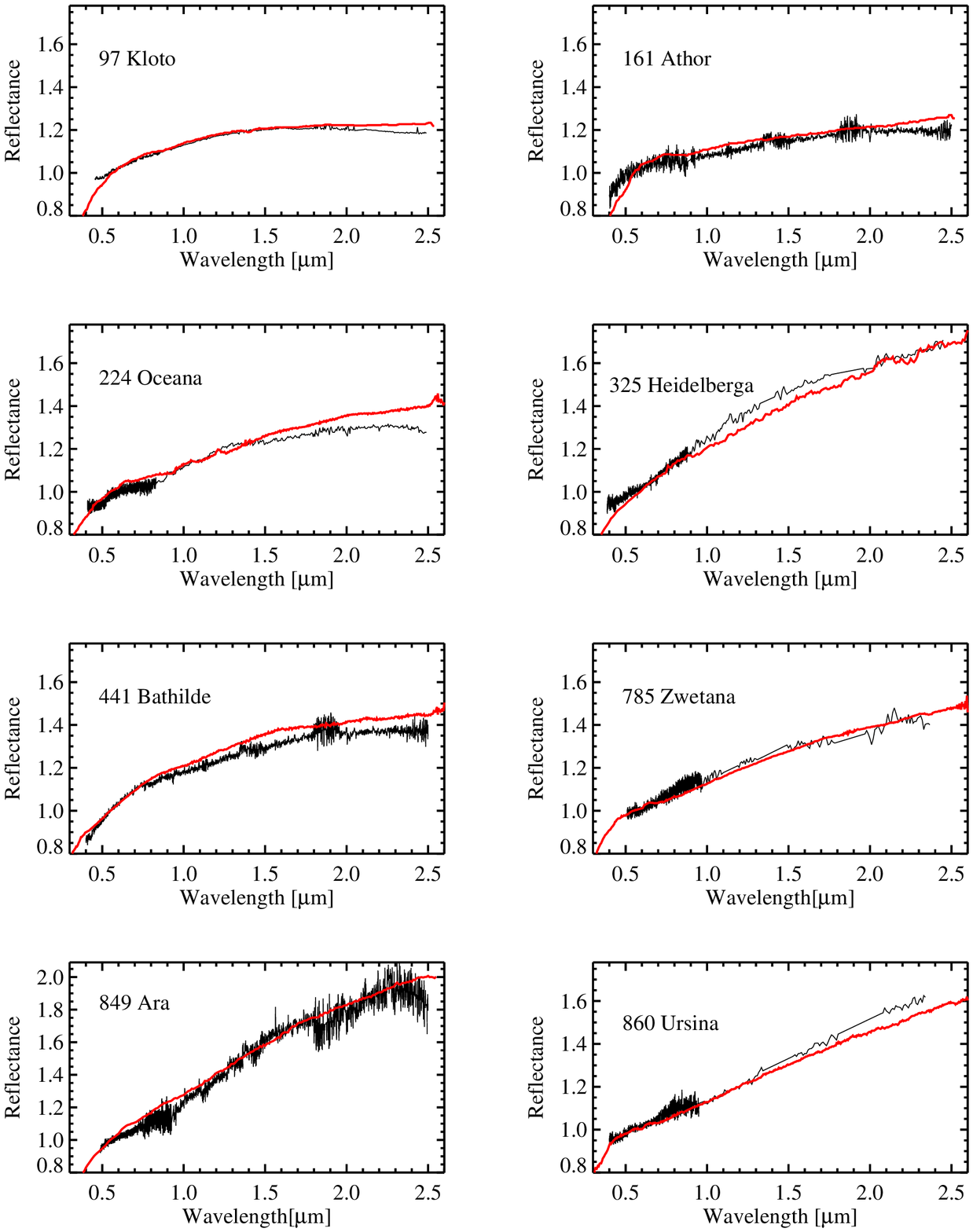,angle=0,width=18.5cm}}
\caption{}
\label{fig6}
\end{figure*}

\begin{figure*}
\centerline{\psfig{file=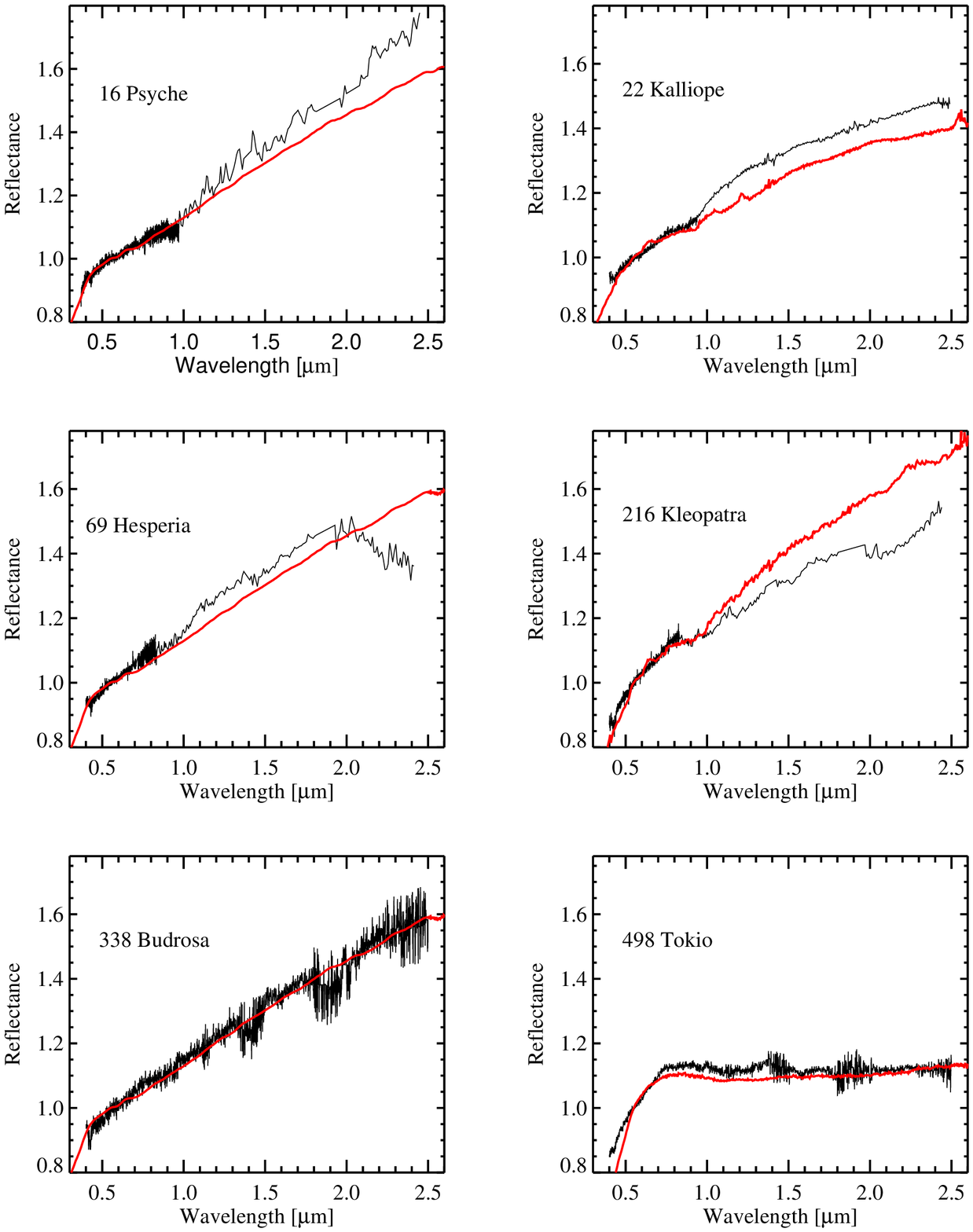,angle=0,width=18.5cm}}
\caption{}
\label{fig7}
\end{figure*}

\begin{figure*}
\centerline{\psfig{file=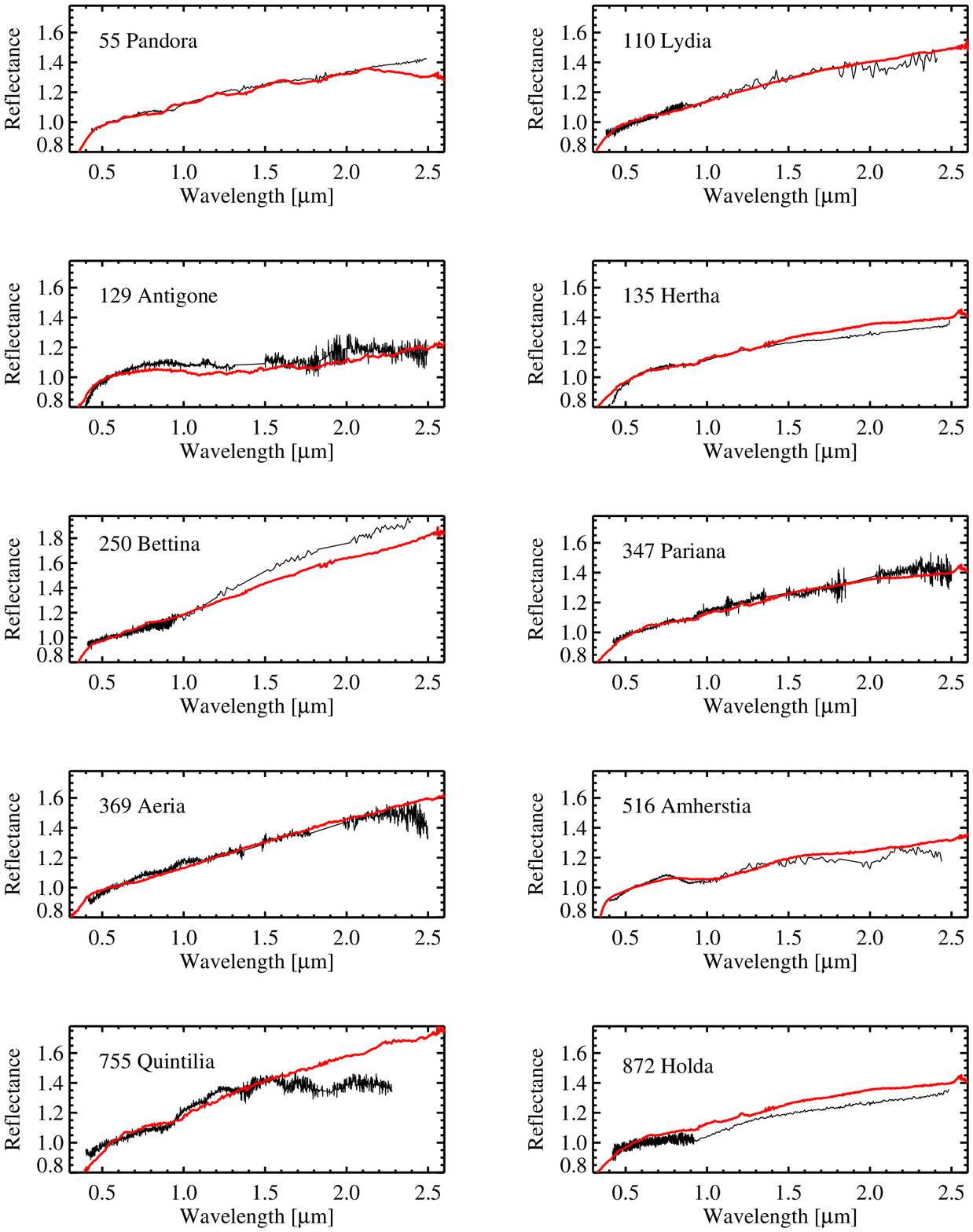,angle=0,width=18.5cm}}
\caption{}
\label{fig8}
\end{figure*}

The best matches between our observed sample and meteorites 
(from the RELAB database) are reported
in Table~\ref{tab} and in Figs~\ref{fig6},~\ref{fig7}, and~\ref{fig8}.
Most best matches resulted in iron meteorites (IM), pallasites (Pall),
or enstatite chondrites (EC), as suggested in the literature for the M-type asteroids. Some meteorite
matches are quite good, while for some asteroids (16 Psyche, 22 Kalliope, 69
Hesperia, 216 Kleopatra 755 Quintilla, 872 Holda) the best meteorite match we
found does not satisfactorily reproduce both the visual and near infrared
spectral behavior. 
The RELAB spectral library does not contain CB meteorite spectra,
so the suggested analogue of CB Bencubbinites (Hardersen et al. 2005)
remains spectrally untested.  In contrast, the RELAB spectral
library contains many samples of other carbonaceous chondrites,
only one of which was found as a best match in this study (498 Tokio was
best fit with a CV carbonaceous chondrite, but its albedo is anomalously low
(7\%) within the M class). This would seem to provide evidence against the
suggestion of Vilas (1994) that carbonaceous
chondrites are analogues for the M-type asteroids. In fact, CV chondrites are
usually associated with K-type asteroids (Moth\'e-Diniz et al. 2008; Clark et
al. 2009), or with B-type asteroids (Clark et al. 2010). 

The only asteroid to result in an enstatite chondrite best match is 161 Athor. 
Neither the asteroid
nor the enstatite chondrites (in general) show absorption features, so the match
is purely based on spectral slope and albedo.

Iron meteorites matched both featureless asteroids (69 Hesperia, 97 Klotho, 224
Oceana, 
325 Heidelberga, 849 Ara) and asteroids with bands 
(16 Psyche, 216 Kleopatra, 110 Lydia, 135 Hertha, 338 Budrosa, 
347 Pariana, 369 Aeria, 860 Ursina, and 872 Holda).
In particular, we note that RELAB file c1mb46 (of the iron meteorite Landes
which
has silicate inclusions) was selected as a best fit for 5 different asteroids
(135 Hertha, 224 Oceana, 347 Pariana, 22 Kalliope, and 872 Holda). 
This spectrum shows a faint absorption band at 0.9 $\mu$m, probably due
to the silicate inclusions.  Of concern, of course, is the fact that
this sample was measured as a slab, not as a particulate.


Pallasites were the best matches for 785 Zwetana (no bands), 110 Lydia, 
250 Bettina, 22 Kalliope, and 55 Pandora (all with bands).  
Several asteroid matches resulted in different RELAB spectral files 
(c1mb43, ckmb43, and cmmb43) of the same meteorite, the stony-iron 
pallasite Esquel.  The differences between these spectra are that 
c1mb43 is of a sanded metal area on an Esquel slab, measured from 
one incidence angle, while cmmb43 is of the same sanded area measured 
from two different incidence angle directions (averaged), and ckmb43 
is a metal-rich powder of the meteorite with a grain size less than 
63 $\mu$m.  In a recent paper, Cloutis et al. (2010) study in detail
the physical reasons for the spectral differences observed as a
function of sample preparation.
This example illustrates a typical issue with a library 
database search.  The same sample can look spectrally different depending 
on measurement illumination and grain size circumstances.  Such spectral 
differences result not from compositional differences, but from 
measurement circumstances.  This indicates that the spectral differences 
between the asteroids found to match these meteorites (110 Lydia, 
22 Kalliope, and 55 Pandora), are insignificant according to the spectral
library.

We note that the peculiar object 516 Amherstia (which we have suggested be
reclassified as an S-type) was found to be similar to a laser-irradiated
ordinary chondrite (OC). 
Similarly, asteroid 441 Bathilde was best fit with a laser-irradiated
sample of olivine.  Unlike 516 Amherstia, 441 Bathilde does not show
absorption features, so the spectral match is based on slopes, shape,
and brightness only.  Nevertheless, we note with amusement that highly
space weathered materials cannot be ruled out as a surface
composition for 441 Bathilde, based
on the excellent agreement between the asteroid and the laboratory spectra
of a simulated weathered olivine.

M-type asteroid spectra, being largely featureless, were very often fit
with a lunar soil sample during the automated part of our RELAB search.
For example, 55 Pandora, which has a band at 0.9 $\mu$m, was well
fit with RELAB file csls65, which is a spectrum
of Apollo program sample 61221,114, a lunar highland soil from the
Apollo 16 landing site.  Like M-type asteroids, lunar soils tend to be 
very steeply spectrally sloped.
When they are most mature (highly space-weathered), lunar soils tend 
to show only very minor silicate absorption features, like M-type asteroids.  
Thus, without applying a "no lunar
sample" filter in our search, we would have been led to the conclusion
that some M-type asteroids are covered with lunar-soil-like materials.

In sum, our method of searching for meteorite analogues emphasized 
the spectral characteristics of brightness and shape, and de-emphasized 
minor absorption bands.  Automated Chi-squared algorithms will always
result in such an emphasis.  We attempted to account for this by selecting
fits with good band mimickry over poor band mimickry when Chi-squared
values were comparable, however there are some cases where weak (1-2\%)
bands are not faithfully matched (e.g. 110 Lydia).
Only spacecraft mission sample returns will ultimately allow us to
answer the question
"Could a 'featureless' iron meteorite be considered an analogue of an asteroid
which has a band?" 

\subsection{Statistics}

[HERE TABLE 6]

[HERE FIGURE 9]

We ran a Spearman Rank Correlation (Spearman 1904) to search for possible
correlations between our sample's spectral characteristics and 
their orbital and physical parameters. The Spearman correlation function gives a
two-element vector containing the rank correlation coefficient ($\rho$) and the 
two-sided significance of its deviation from zero ($P_{r}$).
The value of $\rho$ varies between -1 and 1; if $|\rho|$ is close to zero, then
there is no correlation and if $|\rho|$ is close to 1, then a correlation
exists. The significance ($P_{r}$) is a value in the interval $0 < P_{r} < 1$. A
small value indicates a significant correlation.
We consider a strong correlation to have $P_{r} < 0.01 $ and $|\rho| > 0.6 $,
and a weak correlation to have $P_{r} < 0.08 $ and $ 0.3 < |\rho| < 0.6 $.  

Our sample includes 29 asteroids observed in the visible range. 
We eliminated 516 Amherstia from all further analysis because spectral properties diverge completely from all others. In fact, it displays strong 0.9 and 1.9 $\mu$m features, putting it into the Sq-class in the Bus-DeMeo system. 23 asteroids were observed both in the
visible and near infrared range. We analyse the whole sample as well as the subsamples we consider most likely to be the most ’metal rich’ (those without the 0.9 $\mu$m absorption band) and those containing silicates, presenting the 0.9 $\mu$m band, and belonging to the Xk type in the Bus-DeMeo taxonomy.

Among our Spearman Rank Correlation test results, we find 
weak anti-correlations between the slopes $S_{cont}, S_{NIR1} $, and
$S_{NIR2}$, 
and the asteroids' rotational period (Table~\ref{stat}). 
Belskaya \& Lagerkvist (1996) found that M--asteroids have a faster mean 
rotational period compared to S and C type asteroids. They interpreted this as
an indication of an 
higher mean density that allows  M--asteroids to survive energetic 
collisions without disruption, which increases their rotational 
angular momentum. One possible explanation is that 
fast rotators have a higher density, so they have a composition 
metal rich, which is characterized by steeper spectral slope. 
The slow M-type rotators may have a smaller mean density and so a 
smaller amount of metals in their composition, which is characterized 
by a less steep spectral slope.
      
\begin{figure*}
\centerline{\psfig{file=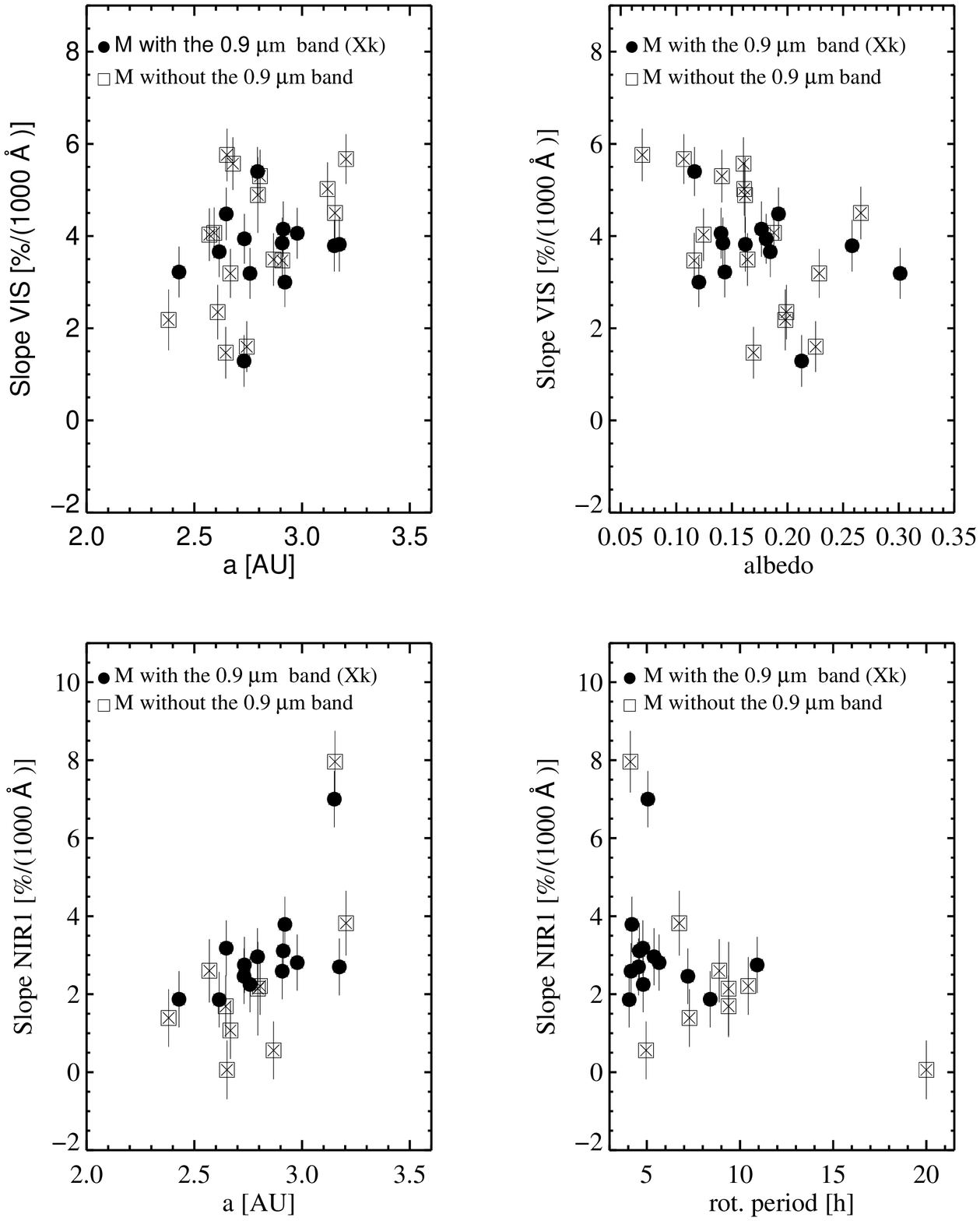,angle=0,width=16.5cm}}
\caption{}
\label{fignn}
\end{figure*}

From our Spearman Rank Correlation tests, there also appear to be 
significant correlations between slope and semi-major axis, and albedo
(Table~\ref{stat}).
The continuum slope ($S_{cont}$) and the 1.1-1.6 $\mu$m slope ($S_{NIR1}$) are
correlated with semi-major axis $a$, while the slope in the visible range ($S_{VIS}$) is only weakly 
correlated with $a$ (Fig.~\ref{fignn}).
Asteroids in our sample with the 0.9 $\mu$m feature show a correlation 
between $S_{cont}$ versus $a$,  and a weaker corelation between $S_{NIR1}$ versus $a$. It must be noted that our sample size is small. 

Considering all 29 asteroids in our sample, a weak anti-correlation exists 
between the $S_{VIS}$ and the albedo, in particular the anti-correlation becomes stronger for asteroids not
showing the silicatic band at 0.9 $\mu$m. So the visual slope tends 
to increase with decreasing albedo.

The correlation of the continuum slope to the semi-major axis together with the
anti correlation of the visible slope to albedo indicates that our sample derived from Tholen M-type asteroids tend to be dark in albedo and red in near-infrared slope with increasing semi-major axis. This is true in particular for 'featureless' subset of our sample that we consider most likely to be iron rich. 
Darkening of albedo plus reddening of the slope is a typical effect of space
weathering seen on lunar samples and silicate rich asteroids. Using statistical
arguments, Lazzarin et al. (2006) have shown that space weathering probably
occurs on all asteroid types.  
Nevertheless, space weathering effects have not been fully investigated for
metallic rich surfaces. If space weathering acts on M-type asteroids as it does
with S type asteroids, then asteroids that are closer to the Sun might
be expected to show stronger spectral changes because of the higher abundance of
solar wind ions. Our results indicate an opposite trend.
A few measurements exist on the effect of space weathering on 
enstatite chondrite and iron and stony iron meteorites
(e.g. Vernazza et al. 2009). An irradiated sample of the Eagle enstatite
chondrite meteorite revealed a slight reddening of the spectral slope  and no
more than 3\% darkening of the visual albedo. For the Vaca Muerta mesosiderite
meteorite Vernazza et al. (2009) found again a reddening in the spectral slope
and a darkening of the visual albedo.
Grain size effects can play an important role in the slope changes as well. 
Britt and Pieters (1988) studied the bidirectional reflectance properties
of some iron-nickel meteorites for different surface roughness. They found that
the spectra of M-type asteroids show good agreement with those of iron
meteorites with surface features in the range of 10 $\mu$m to 1 mm, that is
larger that the wavelength of incident light. Meteorites with these roughness
values are diffuse reflectors and show the classic red slope continuum of iron,
with pratically no geometric dependence on reflection. On the other hand, a  
decreasing of the meteorite surface roughness changes the reflectance
characteristics:  complex scattering behaviour is seen for roughness in the
0.7-10 $\mu$m range, while for roughness values $<$ 0.7 $\mu$m the reflectance
is characterised by two distinct components, the specular one which is bright and
red sloped, and the nonspecular one which is dark and flat. Moreover, Cloutis et
al. (1990) found that the spectra of three iron meteorites reddened with
increasing grain size.  \\
Thus, the correlation of the continuum slope to the semimajor axis together with
the anti correlation of near-infrared slope to albedo are not easily
interpreted. They may reflect a different surface composition of different
M-type asteroids, and/or a combination of surface composition, grain size and
space weathering effects that are not yet fully understood. \\

\section{Discussion}

There are 30 asteroids having M-type visible wavelength and albedo properties that were investigated in this study. One was
removed due to its similarity to an S-type asteroid (516 Amherstia). Of the
remaining 29 asteroids, 16 have detectable faint absorption features. The shape
and slopes in the continuum, visible, and near infrared varies greatly when Tholen M-class objects have their spectral measurements extended.  All were found to be classifiable in the extended wavelength taxonomy defined by DeMeo et al. (2009). Our search for meteorite analogs concluded that the
shape of the spectra could be modeled by a variety of iron and stony iron
meteorites, as well as pallasites, enstatite chondrites, and irradiated
mixtures. 
However, this method of search de-emphasizes the small absorption features. To
investigate
the significance of the shallow absorption features found in our sample, we
used a linear (geographical) spectral mixing model.

One possible avenue for investigation is the mixture of the 
totally featureless, red-sloped metallic meteorites with 
orthopyroxenes, which are thought to create the weak 0.9 $\mu$m feature. 
To constrain the surface composition of our sampled asteroids and the abundance of silicate material needed to reproduce 
the weak 0.9 $\mu$m band seen on some spectra, we created 
spatially segregated (geographical) mixtures of several 
terrestrial and meteoritic materials in different grain sizes.
We considered endmembers from among all the samples 
included in the US Geological Digital Spectral Library 
$http://speclab.cr.usgs.gov/spectral-lib.html$) and in the 
RELAB database. The synthetic spectra that we created were 
compared with the asteroid spectra using the known IRAS 
albedo and the V+NIR spectral behavior and slope as constraints.
A few percent (less than 2\%) of orthopyroxenes or hydrated 
silicates (goethite) added to iron meteorites allowed us to 
reproduce the weak spectral feature around 0.9 $\mu$m. 
Unfortunately, the synthetic spectra did not match both albedo, 
spectral slope, and band depth and center of the observed asteroids. 
Moreover, synthetic spectra do not give unique results, as we 
can obtain similar spectra by adding more components or 
changing the grain size properties of minerals.

[HERE FIGURE 10]

\begin{figure*}
\centerline{\psfig{file=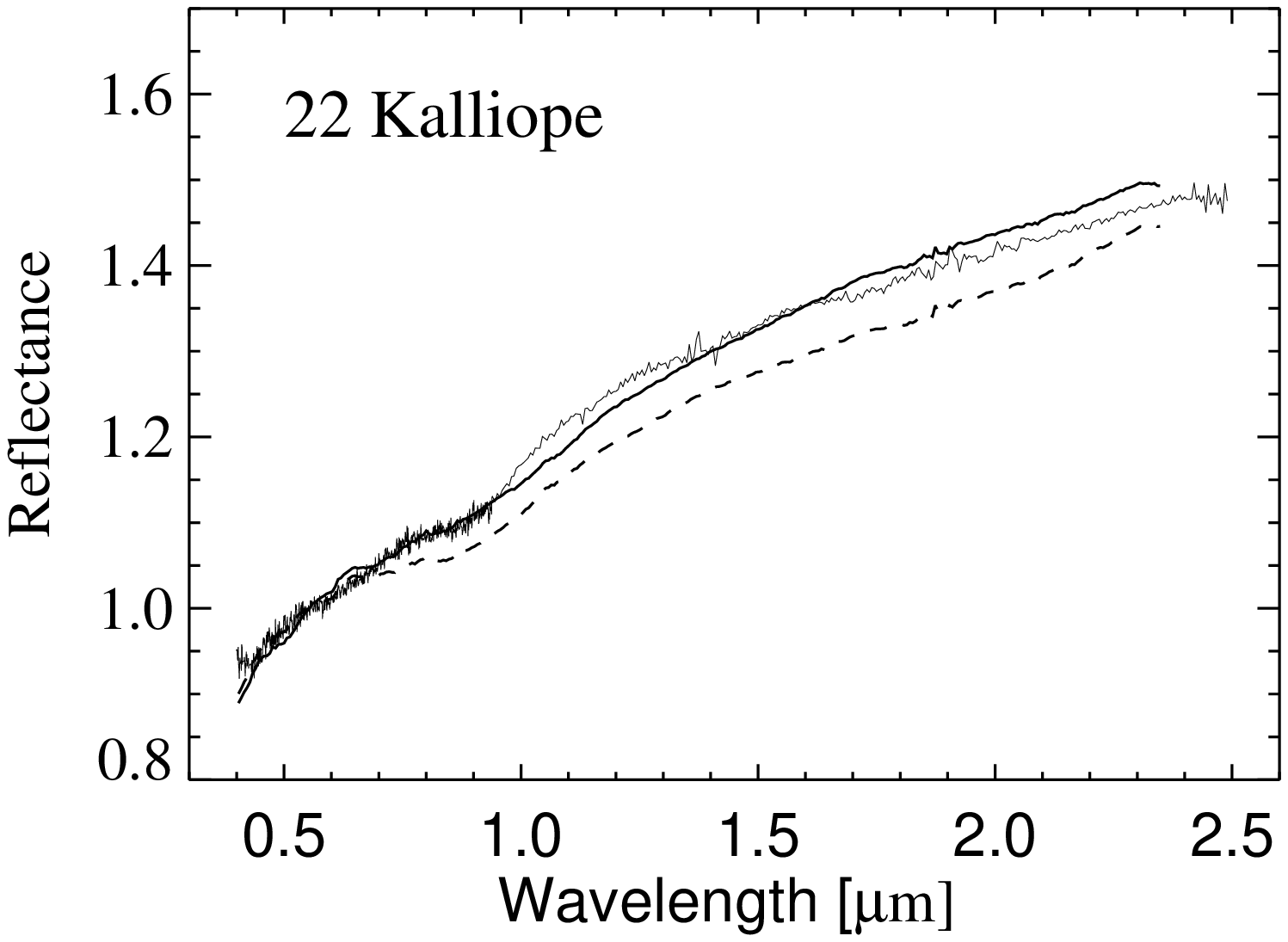,angle=0,width=14cm}}
\caption{}
\label{mixture}
\end{figure*}

In Fig.~\ref{mixture} we show an example of two different 
mixtures proposed for the asteroid 22 Kalliope. 
The solid line represents a mixture with 98\% pallasite 
(RELAB file ckmb43) and 2\% goethite, the dashed line shows 
a mixture of the same meteorite enriched with 1\% of 
orthopyroxene (grain size of 25--45 $\mu$m).  The spectral 
behaviour of both mixtures does not completely match all 
the spectral characteristics of 22 Kalliope. Figure~\ref{mixture}
emphasizes our results that a) it is not easy to fully reproduce 
the 0.9 $\mu$m band, asteroid albedo and spectral slope simply 
enriching an iron meteorite with some silicates, b) 
the 0.9 $\mu$m band could be reproduced both with a small 
amount of anhydrous silicates such as orthopyroxene, but also 
with some hydrated silicates like goethite. This last 
possibility cannot be completely ruled out. It would 
support the hypothesis of the presence of hydrated materials on 
some M-type asteroids, as inferred by Rivkin et al. (1995; 2000) 
with the detection of the 3 $\mu$m band.

Cloutis et al. (1990) identified the 0.9 $\mu$m feature 
in the spectra of some enstatite chondrites. However, 
the shape of the spectra in our study is similar 
to iron meteorites, which are featureless. 
Feierberg et al. (1982) have shown that the physical 
state of the surface plays an important role in the 
relative contribution of metal to the spectrum
of a silicate-metal mixture. If the surface is spotty 
with separate areas of pure silicate and pure metal, 
the spectrum differs very little from that of pure silicate, 
even in the case of 50\% metal content. But if metal 
and pyroxene particles were mixed, the presence of 50\% metal almost
completely suppresses the pyroxene bands and the 
spectrum appears to be indistinguishable
from that of the pure metal. Thus, our spectra could 
have a fairly high concentration of silicate materials, 
as long as the surface composition is well-mixed.

The results of our search for meteorite analogs confirm that these objects originally in the Tholen M-class could be the parent 
bodies both of iron (or stony-iron) meteorites and enstatite chondrites. 
Note that the enstatite chondrites were supposed to have formed 
close to the Sun (1-1.4 AU) (Shukolyukov and Lugmair 2004). 
Together with aubrites, enstatite chondrites are the only 
group of meteorites that have
the same oxygen isotopic composition as the Earth--Moon 
system (Clayton et al. 1984). It is excluded that the 
parent bodies of enstatite chondrites still reside so close 
to the Sun,  but they may have been scattered into the 
main belt ($\sim$ 2.2-3.0 AU), as described by the 
Bottke et al.(2006) migration scenario. The same 
hypothesis has been proposed by Bottke et al. (2006) 
for iron meteorites, that is iron meteorite parent bodies 
that formed in the terrestrial planet region with 
fast accretion times. This allowed small planetesimals to 
melt early in Solar System history by the decay of short-lived 
radionuclides, then be scattered in the main belt.

\section{Conclusions}

We observed and analyzed spectra obtained in the visible and near 
infrared wavelength range (0.4-2.5 $\mu$m) of 30 asteroids selected according to their original classifications as M-types, following the Tholen (1984) taxonomy. The results of our investigation are the following:
\begin{itemize}
\item  Tholen M-type asteroids show a large variety of spectral behaviors when their spectral coverage is extended into the near-infrared.
\item Several weak absorption bands have been identified in extending the spectral range for objects defined as Tholen M-types, showing that their surface compositions are not exclusively 
metallic. In particular on several asteroids we identify 
a weak feature (1-5 \% depth) in the 0.9 $\mu$m wavelength 
region attributed to low-Fe, low-Ca orthopyroxene minerals. 
All these asteroids belong to the Xk class in the Bus-DeMeo taxonomy. 
Some asteroids show a band at 0.43 $\mu$m (16 Psyche, 22 
Kalliope, 69 Hesperia, 216 Kleopatra, 338 Budrosa,
and 498 Tokio) that can be attributed to chlorites 
and Mg-rich serpentines, or to pigeonite or augite, or 
to an Fe$^{3+}$  spin-forbidden transition in the iron 
sulfate jarosite. 132 Aethra shows a 0.49 $\mu$m 
band typical of the Xe type (following the Bus\& Binzel 
(2002b) taxonomy) probably due to sulfides such as 
oldhamite and/or troilite. 135 Hertha has a faint 
band at $\sim$ 0.51 $\mu$m similar to the Fe$^{2+}$ 
spin-forbidden crystal field transitions in Earth and lunar pyroxene. 
The spectrum of 755 Quintilla is peculiar with four 
absorption bands:  0.9 and 1.9 $\mu$m features could be 
attributed to pyroxene silicates, but the origins of the 
1.37 and 1.61 $\mu$m bands are unknown.
\item A comparison with the published literature reveals 
that at least 4 asteroids (129 Antigone, 325 Heidelberga, 498 Tokio, and 785 Zwetana) display surface variability. Seven additional objects here investigated (16 Psyche, 201 Penelope, 250 Bettina, 369 Aeria, 516 Amherstia, 558 Carmen, and 755 Quintilla) have slightly different spectral slope values or band center positions as compared to the data already published in the literature. These variations may be real but we cannot exclude that they are simply due to the observational uncertainties or to different data acquisition and reduction processes.
\item We performed a search of meteorite and/or 
mineral spectral matches between our asteroid sample and the RELAB database. We confirm that these objects could be 
the parent bodies both of iron (or stony-iron) meteorites and 
enstatite chondrites, as suggested in the literature. 
Only 498 Tokio, which has an anomalous low albedo value (7\%) within the Tholen definition for the M-type population, is best matched by a CV carbonaceous chondrite.
\item Our attempt to model the asteroid spectra with 
mixtures of iron/stony iron meteorites enriched with a 
few percent silicates did not match the albedo and spectral 
characteristics of our sample. The amount of silicates 
needed to reproduce the weak 0.9 $\mu$m feature seen on 
some asteroids is very small, less than 2\%.
\item A statistical analysis of our sampled asteroids'
spectral characteristics versus orbital elements and 
physical parameters indicated a possible correlation 
between the continuum slope and semi-major axis, and 
two anti-correlations between visible slope to albedo and 
continuum slope to rotational period. 
Our asteroid sample derived from the Tholen M-type 
tends to be dark in albedo and red in near infrared slope 
with increasing value of the semi-major axis. 
Fast M-type rotators tend to be redder than slow rotators.

\end{itemize}
\bigskip

{\bf Acknowledgments} \\
We thank Prof R. Binzel and an anonymous referee for their very useful comments
that help us improving the paper. 
We thank Dr. F. DeMeo for her help in the classification of the observed 
asteroids in her taxonomy. Taxonomic-type results presented in this work 
were determined in part using a Bus-DeMeo Taxonomy Classification 
Web tool developed by Stephen M. Slivan at MIT with the support of 
National Science Foundation Grant 0506716 and NASA Grant NAG5-12355. 
This research utilizes spectra acquired with
the NASA RELAB facility at Brown University. This paper is in part based on 
observations made with the Italian Telescopio
Nazionale Galileo (TNG) operated on the island of La Palma by 
the Centro Galileo Galilei of the INAF (Istituto Nazionale di
Astrofisica) at the Spanish Observatorio del Roque de los Muchachos 
of the Instituto de Astrofisica de Canarias. BEC thanks the 
Observatoire de Paris for kind hospitality during her sabbatic
from Ithaca College in the Spring of 2009.

\bigskip

{\bf References} \\
Adams, J. B. 1975. Interpretation of visible and near-infrared reflectance
spectra of pyroxenes other rock-forming minerals. In Infrared and Raman
Spectroscopy of Lunar and Terrestrial Minerals (C. Karr, Jr., Ed.), pp. 90–116.
Academic Press, New York

Baffa, C., Comoretto, G., Gennari, S., Lisi, F., Oliva, E., Biliotti, V.,
Checcucci, A., Gavrioussev, V., Giani, E., Ghinassi, F., Hunt, L. K., Maiolino,
R., Mannucci, F., Marcucci, G., Sozzi, M., Stefanini, P., Testi, L., 2001. NICS:
The TNG Near Infrared Camera Spectrometer. Astron. Astroph. 378, 722-728

Bell, J.F., Davis, D., Hartmann, W.K., Gaffey, M.J., 1989. Asteroids: The big
picture. In: Binzel, R.P., Gehrels, T., Matthews, M.S. (Eds.), Asteroids II.
Univ. of Arizona Press, Tucson, pp. 921–948

Belskaya, I. N., Lagerkvist, C.I., 1996. Physical properties of M class
asteroids. Planet. Space Sci. 44,  783--794

Birlan, M., Vernazza, P., Nedelcu, D. A., 2007. Spectral properties of nine
M-type asteroids. Astron. Astroph. 475, 747--754

Binzel, R. P., Bus, S. J., Sunshine, J., and Burbine, T. H., 1995. Rotationally Resolved Spectra of Asteroid 16 Psyche. Icarus  117, 443-445

Bottke, W.F., Nesvorný, D., Grimm, R.E., Morbidelli, A., O’Brien, D.P., 2006.
Iron meteorites
as remnants of planetesimals formed in the terrestrial planet region. Nature
439, 821–824

Britt, D.T.,Pieters, C. M., 1988. Bidirectional reflectance properties of
iron-nickel meteorites. In: Lunar and Planetary Science Conference, 503-512.

Britt, D.T., Yeomans, D., Housen, K., Consolmagno, G., 2002. Asteroid density,
porosity, and structure. In: Bottke Jr.,W.F., Cellino, A., Paolicchi, P.,
Binzel,
R.P. (Eds.), Asteroids III. Univ. of Arizona, Tucson, pp. 485–500

Burbine, T. H., Cloutis, E. A., Bus, S. J., Meibom, A., Binzel, R. P., 1998. The
detection of troilite (FeS) on the surfaces of E-class asteroids.  Bull. Am.
Astron. Soc. 30, 711
 
 Burbine, T. H., McCoy, T. J., Nittler, L., Benedix, G., Cloutis, E., Dickenson,
T. 2002a. Spectra of extremely reduced assemblages: Implications for Mercury.
Meteorit. Planet. Sci. 37, 1233–1244

 Burbine, T.H., McCoy, T.J., Meibom, A., 2002b. Meteorite parent bodies, in
Asteroids III, Bottke, W. et al.
Editors, Univ. of Arizona Press, Tucson.

Burns, R. G., D. J. Vaughan, R. M. Abu-Eid, M. Witner, and A. Morawski 1973.
Spectral evidence for Cr$^{3+}$, Ti$^{3+}$, and Fe$^{2+}$ rather than Cr$^{2+}$,
and Fe31 in lunar ferromagnesian silicates. In Proc. 4th Lunar Sci. Conf.
983–994

 Bus, S. J., Binzel, R. P.,  2002a. Phase II of the Small Main-Belt Asteroid
Spectroscopic Survey: The Observations. Icarus 158, 106-145

  Bus, S. J., Binzel, R. P.,  2002b. Phase II of the Small Main-Belt Asteroid
Spectroscopic Survey: A Feature-Based Taxonomy. Icarus 158, 146-177 

Bus, S.J., Binzel, R.P., Volquardsen, E., 2003. Characterizing the visible and
near-IR spectra of asteroids using principal component analysis. Bull. Amer.
Astron. Soc. 35, 976 (abstract)
	
Busarev, V. V., 1998. Spectral Features of M-Asteroids: 75 Eurydike and 201
Penelope. Icarus, 131, 32--40

Chapman, C. R., Gaffey, M. J., 1979. Spectral reflectances of the asteroids. In
Asteroids (Gehrels, Ed.), pp. 1064-1089. Univ. of Arizona Press, Tucson

Clayton, R.N., Mayeda, T.K., Rubin, A.E., 1984. Oxygen isotopic compositions of
enstatite
chondrites and aubrites. In: Proc. 15th Lunar Planet. Sci. Conf. J. Geophys. Res. Suppl. 89, C245–C249

Clark, B. E., Bus, S. J., Rivkin, A. S., Shepard, M. K., Shah, S., 2004a.
Spectroscopy of X-Type Asteroids. Astron. J 128, 3070-3081

Clark, B.E., Bus, S.J., Rivkin, A.S., McConnochie, T.,
Sander, J., Shah, S., Hiroi, T., Shepard, M., 2004b. E-Type asteroid
spectroscopy and compositional modeling, JGR 109, 1010-1029

Clark, B.E., Ockert-Bell, M.E., Cloutis, E.A., Nesvorny, D., Mothé-Diniz, T.,
and Bus, S.J. 2009. Spectroscopy of K-complex Asteroids: Parent Bodies of
Carbonaceous Meteorites?  Icarus 202, 119-133

Clark, B.E., Ziffer, J., Nesvorny, D., Campins, H., Rivkin, A.S., Hiroi, T.,
Barucci, M.A., Fulchignoni,
M., Binzel, R.P., Fornasier, S., DeMeo, F., Ockert-Bell, M.E., Licandro, J., and
Mothé-Diniz, T., 2010.
Spectroscopy of B-type Asteroids: Subgroups and Meteorite Analogs.  
JGR Planets, in press

Cloutis, E. A., Gaffey, M. J., Smith, D. G. W., Lambert, R. St. J., 1990. Metal
Silicate Mixtures: Spectral Properties and Applications to Asteroid Taxonomy. J.
Geophys. Res., 95, 281, 8323–8338

Cloutis, E. A., and T. H. Burbine 1999. The spectral properties of
troilite/pyrrhotite and implication
s for the E-asteroids. {\it Lunar and Planet. Sci. Conf. XXX}, 1875

Cloutis, E.A., Hardersen, P.S., Bish, D.L., Bailey, D.T., Gaffey, M.J., Craig, M.A., 2010. Reflectance spectra of iron meteorites: Implications for spectral identification of their parent bodies. Meteoritics and Planetary Science, in press.

Descamps, P., Marchis, F., Pollock, J., Berthier, J., Vachier, F., Birlan, M.,
Kaasalainen, M., Harris, A. W., Wong, M. H., Romanishin, W. J., Cooper, E. M.,
Kettner, K. A., Wiggins, P., Kryszczynska, A., Polinska, M., Coliac, J.F.,
Devyatkin, A., Verestchagina, I., Gorshanov, D., 2008. New determination of the
size and bulk density of the binary Asteroid 22 Kalliope from observations of
mutual eclipses. Icarus 196, 578-600

DeMeo, F. E., Binzel, R P., Slivan, S. M., Bus, S. J., 2009. An extension of the
Bus asteroid taxonomy into the near-infrared. Icarus 202, 160-180

D'Odorico, S., 1990. EMMI, the ESO multi-mode instrument, successfully installed
at the NTT.The Messenger 61,  51-56

Feierberg, M. A.,  Larson, H.P., Chapman, C. R., 1982. Spectroscopic evidence
for undifferentiated
S-type asteroids. Astrophys. J. 257, 361-372
	
Fieber-Beyer, S. K., Gaffey, M. J., Hardersen, P. S., 2006. Near-Infrared
Spectroscopic Analysis of Mainbelt M-Asteroid 755 Quintilla. 37th Annual Lunar
and Planetary Science Conference, March 13-17, 2006, League City, Texas,
abstract no.1315

Fornasier, S., Dotto, E., Marzari, F., Barucci, M.A., Boehnhardt, H., Hainaut,
O., de Bergh, C., 2004a. Visible spectroscopic and photometric survey of L5 Trojans
: investigation of dynamical families. Icarus 172,  221--232

 Fornasier, S., Dotto, E., Barucci, M. A., Barbieri, C., 2004b. Water ice on the
surface of the large TNO 2004 DW. Astron. Astroph. 422, L43-L46
 
Fornasier, S., Marzari, F., Dotto, E., Barucci, M. A., Migliorini, A., 2007.
Are the E-type asteroids (2867) Steins, a target of the Rosetta mission,
and NEA (3103) Eger remnants of an old asteroid family? Astron. Astroph. 474,
29-32
	
Fornasier, S., Migliorini, A., Dotto, E., Barucci, M. A., 2008. Visible and near
infrared spectroscopic investigation of E-type asteroids, including 2867 Steins,
a target of the Rosetta mission. Icarus 196, 119--134

Gaffey, M. J., Burbine, T. H., Piatek, J. L., Reed, K. L., Chaky, D. A., Bell J.
F. Brown, R. H., 1993. Mineralogical variations within the S-type asteroid
class.
Icarus 106, 573--602

Gaffey,  M. J., 1976. Spectral reflectance characteristics of the meteorite
classes. J. Geophys. Res. 81, 905--920
	
Hardersen, P. S., Gaffey, M. J., Abell, P. A., 2005. Near-IR spectral evidence
for the presence of iron-poor orthopyroxenes on the surfaces of six M-type
asteroids. Icarus 175, 141--158

Hardersen, P. S., Gaffey, M. J., Cloutis, E., Abell, P. A., Reddy, V., 2006a.
Discovering Spectral and Mineralogical Diversity Among the M-Asteroid
Population. 37th Annual Lunar and Planetary Science Conference, 2006, League
City, Texas, abstract no. 1106

Hardersen, P. S., Gaffey, M. J., Abell, P. A., 2006b. Near-infrared Reflectance
Spectra Of 135 Hertha, 224 Oceana, 516 Amherstia, And 872 Holda. Bulletin of the
American Astronomical Society, Vol. 38, p.626

Hardersen, P. S., Gaffey, M. J., Kumar, S., 2007a. Nir Spectra And
Interpretations For M-asteroids 369 Aeria And 785 Zwetana. Bulletin of the
American Astronomical Society, Vol. 39, p . 478

Hardersen, P. S., Gaffey, M. J., Kumar, S., Fieber-Beyer, S. K., Crowell, J. J.,
Crowell, A. M., 2007b. Near-IR Reflectance Spectra of M-Asteroids 250 Bettina,
369 Aeria, 413 Edburga, and 931 Whittemora. 38th Lunar and Planetary Science
Conference, March 12-16, 2007 in League City, Texas. LPI Contribution No. 1338,
p.1956
	
Hazen, R. M., P. M. Bell, and H. K. Mao 1978. Effects of compositional variation
on absorption spectra of lunar pyroxenes. In Proc. 9th Lunar Planet. Sci. Conf.,
2919–2934.

Jones, T. D., Lebofsky, L. A., Lewis, J. S.,  Marley, M. S., 1990. 
The composition and origin of the C, P, and D asteroids: Water
as a tracer of thermal evolution in the outer belt. Icarus, 88, 172–192

King, T.V.V., Clark, R.N., 1989. Spectral characteristics of chlorites and
Mgserpentines
using high resolution reflectance spectroscopy. J. Geophys. Res. 94,
13997–14008

Kochetova, O.M., 2003. Application of new criteria for selecting perturbed minor
planets for determination of masses of perturbing minor planets by
dynamical method. In: Reports of Inst. of Applied Astronomy of Russian
Acad. Sci. No. 165, St. Petersburg, 42 pp. (in Russian).
	
Lazzarin, M., Marchi, S., Moroz, L. V., Brunetto, R., Magrin, S., Paolicchi, P.,
Strazzulla, G., 2006.
Space Weathering in the Main Asteroid Belt: The Big Picture. Astrophysical J.
647, 179-182

Lord, S., 1992. A New Software Tool for Computing Earth's Atmospheric
Transmission of Near- and Far-Infrared Radiation. NASA Technical Memorandum
103957

Lupisko, D. F., Belskaya, I. N., 1989. On the surface composition of the M-type
asteroids. Icarus 78, 395--401

Magri, C. Nolan, M. C.; Ostro, S. J., Giorgini, J. D., 2007. A radar survey of
main-belt asteroids: Arecibo observations of 55 objects during 1999 2003. Icarus
186, 126--151

Margot, J.-L., Brown, M.E., 2003. A low-density M-type asteroid in the mainbelt.
Science 300, 1939–1942

Moorwood, A., Cuby, J.-G., Lidman, C., 1998. SOFI sees first light at the NTT.
The Messenger 91, 9-13

Mothé-Diniz, T., J.M. Carvano, S.J. Bus, R. Duffard, T. Burbine, 2008.
Mineralogical
analysis of the Eos family from near-infrared spectra.  Icarus 195, 277-294

Ockert-Bell, M.E., Clark, B.E., Shepard, M.K., Rivkin, A.S., Binzel, R.P.,
Thomas, C.A., DeMeo, F.E., Bus, S.J., Shah, S., 2008. Observations of X/M
asteroids across multiple wavelengths. Icarus 195 (1), 206–219

Ostro, S. J., Campbell, D. B., Chandler, J. F., Hine, A. A., Hudson,
R. S., Rosema, K. D., Shapiro, I. I., 1991. Asteroid 1986
DA: Radar evidence for a metallic composition. Science, 252,
1399--1404

Ostro, S. J., Hudson, R. S., Nolan, M. C., Margot, J. L., Scheeres,
D. J., Campbell, D. B., Magri, C., Giorgini, J. D., Yeomans,
D. K., 2000. Radar observations of asteroid 216 Kleopatra.
Science, 288, 836–839

Pieters, C. 1983. Strength of mineral absorption features in the transmitted
component of near-infrared reflected light: First results from RELAB, J.
Geophys. Res., 88, 9534--9544

Rayner, J.T., Toomey, D., Onaka, P., Denault, A., Stahlberger, W., Vacca, W.,
Cushing, M., 2003. SpeX: A Medium-Resolution 0.8-5.5 μm Spectrograph and Imager
for the NASA Infrared Telescope Facility. Pub. Astron. Soc. Pacific, 115 (805),
362-382

Rivkin, A. S., Howell, E. S., Britt, D. T., Lebofsky, L. A., Nolan,
M. C., Branston, D. D., 1995. 3-μm spectrophotometric
survey of M- and E-class asteroids. Icarus, 117, 90--100

Rivkin, A.S., Howell, E.S., Lebofsky, L.A., Clark, B.E., Britt, D.T., 2000.
The nature of M-class asteroids from 3-μm observations. Icarus 145, 351--368

Rivkin, A. S., Howell, E. S., Vilas, F., Lebofsky L. A., 2002.
Hydrated minerals on asteroids: The astronomical record. In
Asteroids III, Bottke W. et al. editors, Univ. of Arizona, Tucson, pp 235--253
	
Shepard, M. K., Clark, B. E., Nolan, M. C., Howell, E. S., Magri, C., Giorgini,
J. D., Benner, L. A. M., Ostro, S. J., Harris, A. W., Warner, B., et al., 2008.
A radar survey of M- and X-class asteroids. Icarus 195, 184--205	
	
Shukolyukov, A., Lugmair, G.W., 2004. Manganese–chromium isotope systematics of
enstatite meteorites. Geochim. Cosmochim. Acta 68, 2875–2888	
	
Spearman, C. 1904. The proof and measurements of associations between two
things. AM. J. Psychol. 57, 72
	
Sunshine, J. M., Shelte, S. J., McCoy, T. J., Burbine, T. H., Corrigan,
C. M., Binzel, R. P., 2004. Highcalcium
pyroxene as an indicator of igneous differentiation in asteroids and meteorites.
Meteoritics and Planetary Sci. 39, 1343-1357

Takir, D., Hardersen, P. S., Gaffey, M. J., 2008. The Near-Infrared Spectroscopy
of Two M-Class Main Belt Asteroids, 77 Frigga and 325 Heidelberga. 39th Lunar
and Planetary Science Conference, League City, Texas. LPI Contribution No.
1391., p.1084

Taylor, S. R., 1992.  Solar system evolution: a new perspective. an inquiry into
the chemical composition, origin, and evolution of the solar system. In Solar
System evolution: A new Perspective, Cambridge Univ. Press.

Tedesco, E.F., P.V. Noah, M. Moah, and S.D. Price, 2002. The supplemental IRAS
minor planet survey.
The Astronomical Journal 123, 10565-10585

Tholen, D.J., 1984. Asteroid taxonomy from cluster analysis of photometry.
Ph.D. dissertation, University of Arizona, Tucson

Tholen, D.J., Barucci, M.A., 1989. Asteroids taxonomy. In: Binzel, R.P.,
Gehrels, T., Matthews, M.S. (Eds.), Asteroids II. Univ. of Arizona Press,
Tucson, pp. 298--315

Vernazza, P., Brunetto, R., Binzel, R. P., Perron, C., Fulvio, D., Strazzulla,
G., Fulchignoni, M., 2009. Plausible parent bodies for enstatite chondrites and
mesosiderites: Implications for Lutetia's fly-by. Icarus 202, 477-486

Viateau, B., 2000. Mass and density of Asteroids (16) Psyche and (121)
Hermione. Astron. Astrophys. 354, 725–731

Vilas, F., Hatch, E.C., Larson, S.M., Sawyer, S.R., Gaffey, M.J., 1993. Ferric
iron in primitive asteroids: a 0.43 μm absorption feature. Icarus 102, 225–
231

Vilas, F., 1994. A cheaper, faster, better way to detect water of hydration on
Solar System bodies. Icarus 111, 456–467

\newpage
\clearpage

{\bf Tables}

{\scriptsize
       \begin{center}
     \begin{longtable} {|l|l|c|c|c|c|c|c|l|}  
\caption[]{Observational circumstances for the observed sample, for which their visible wavelength spectra and albedos all satisfy the definition of M-type in the Tholen taxonomy. Solar analog stars named "hip" come from the Hipparcos catalogue, "la" from the
Landolt photometric standard stars catalogue, and "HD" from the Henry Draper
catalogue}. 
        \label{tab1} \\
\hline \multicolumn{1}{|c|} {\textbf{Object    }} & \multicolumn{1}{c|}
{\textbf{Night}} & \multicolumn{1}{c|} {\textbf{UT$_{start}$}} &
\multicolumn{1}{c|} {\textbf{T$_{exp}$}} &     \multicolumn{1}{c|}
{\textbf{Tel.}} & \multicolumn{1}{c|} {\textbf{Instr.}} & \multicolumn{1}{c|}
{\textbf{Grism}} & \multicolumn{1}{c|} {\textbf{airm.}} & \multicolumn{1}{c|}
{\textbf{Solar Analog (airm.)}} \\  \hline 
\endfirsthead
\multicolumn{9}{c}%
{{\bfseries \tablename\ \thetable{} -- continued from previous page}} \\ \hline 
\endfoot
\hline \multicolumn{1}{|c|} {\textbf{Object    }} & \multicolumn{1}{c|}
{\textbf{Night}} & \multicolumn{1}{c|} {\textbf{UT$_{start}$}} &
\multicolumn{1}{c|} {\textbf{T$_{exp}$}} &     \multicolumn{1}{c|}
{\textbf{Tel.}} & \multicolumn{1}{c|} {\textbf{Instr.}} & \multicolumn{1}{c|}
{\textbf{Grism}} & \multicolumn{1}{c|} {\textbf{airm.}} & \multicolumn{1}{c|}
{\textbf{Solar Analog (airm.)}} \\  \hline 
\endhead
\hline \multicolumn{9}{r}{{Continued on next page}} \\ 
\endfoot
\hline \hline
\endlastfoot
16 Psyche     & 15 Nov. 04 & 19:31 & 40 & TNG & Dolores & LR-R & 1.46 & la115-271 (1.31) \\
16 Psyche     & 15 Nov. 04 & 19:33 & 60 & TNG & Dolores & MR-B & 1.47 & la115-271 (1.32) \\
16 Psyche     & 18 Nov. 04 & 19:21 & 80 & TNG & NICS & Amici   & 1.45  & la93-101 (1.34) \\
22 Kalliope   &13 Aug. 05&07:35	&120& NTT & EMMI &	GR1 &	1.02  & HD1835 (1.07) \\
22 Kalliope   &21 Dec. 07&14:46 & 1800 & IRTF & SPEX & Prism  & 1.24 & hyades64(1.00),la115-270(1.10), \\
& & & & & & & & la93-101(1.10), la102-1081(1.10)\\
55 Pandora  & 25 May 04  & 10:33 & 60 & NTT  & EMMI    & GR1  & 1.31 & la112-1333 (1.17) \\
55 Pandora  & 17 Sep. 05 & 11:37 & 1920 &   IRTF & SPEX & Prism&	1.10  & hyades64 (1.00),la115-270(1.10), \\
& & & & & & & & la93-101(1.10), la112-1333(1.10) \\
55 Pandora  & 20 Sep. 05 & 13:56 & 1000 &	 IRTF &	SPEX &Prism &1.00 & la110-361(1.10),la115-270(1.10) \\
& & & & & & & & ,la93-101(1.10), la112-1333(1.10) \\
69 Hesperia & 18 Nov. 04 & 06:34 & 120 &  TNG & Dolores & LR-R & 1.12 & hyades64 (1.4) \\ 
69 Hesperia & 18 Nov. 04 & 06:37 & 180 &  TNG & Dolores & MR-B & 1.13 & hyades64 (1.4) \\ 
69 Hesperia & 19 Nov. 04 & 06:01 & 480 & TNG & NICS & Amici   & 1.16  & la98-978 (1.21) \\
97 Klotho      &  20 Jan 07 & 03:12 & 20 & NTT & EMMI &	GR1 & 1.24 & la98-978 (1.22) \\
97 Klotho     &	20 Sep. 05 &06:06 & 960 & IRTF &	SPEX &	Prism &	1.20 & la110-361(1.10),la115-270(1.10), \\
& & & & & & & & la93-101(1.10), la112-1333(1.10) \\
97 Klotho     &	17 Dec 06 &06:06  & 670  &	IRTF &	SPEX &	Prism &	1.1 & hyades64 (1.15) \\
97 Klotho     &	19 Dec 06 &10:48  & 600 &	IRTF &	SPEX &	Prism &	1.1 & hyades64 (1.15), la102-1081 (1.1)\\
110 Lydia & 16 Nov. 04 & 05:24 & 180 &  TNG & Dolores & LR-R & 1.12 &   HD28099 (1.05) \\
110 Lydia & 16 Nov. 04 & 05:28 & 180 &  TNG & Dolores & MR-B & 1.11 &   HD28099 (1.05) \\
110 Lydia & 19 Nov. 04 & 05:03 & 360 & TNG & NICS & Amici   & 1.14  & la98-978 (1.17) \\
125 Liberatrix & 20 Nov. 04 & 20:37 & 180 & TNG & Dolores & LR-R  & 1.22 & la115-271 (1.16)      \\
125 Liberatrix & 20 Nov. 04 & 20:41 & 180 & TNG & Dolores & MR-B  & 1.21 & la115-271 (1.16)   \\
125 Liberatrix & 20 Nov. 04 & 22:01 & 150 & TNG & Dolores & LR-R  & 1.12 & la115-271 (1.16)      \\
125 Liberatrix & 20 Nov. 04 & 22:05 & 150 & TNG & Dolores & MR-B  & 1.12 & la115-271 (1.16)   \\
125 Liberatrix & 20 Nov. 04 & 23:27 & 180 & TNG & Dolores & LR-R  & 1.18 & la115-271 (1.16)      \\
125 Liberatrix & 20 Nov. 04 & 23:31 & 180 & TNG & Dolores & MR-B  & 1.19 & la115-271 (1.16)   \\
129 Antigone  &13 Aug 05 & 00:07 & 60 & NTT & EMMI & GR1  &1.16 & la112-1333 (1.19) \\        	
129 Antigone  &14 Aug 05 & 02:21 & 160 & NTT & SOFI & GBF  &1.81  & la107-684 (2.03) \\
129 Antigone  &14 Aug 05 & 02:25 & 200 & NTT & SOFI & GRF  &1.85  & la107-684 (2.03) \\	
132  Aertha   &25 May 04 & 10:45 & 180 & NTT & EMMI & GR1  &1.43 & la112-1333 (1.17) \\
135 Hertha & 26 May 04  & 09:58 & 120 & NTT  & EMMI    & GR1 & 1.27 &  la112-1333 (1.18) \\
135 Hertha & 07 Nov. 08	& 10:28	& 630 &	IRTF &	SPEX &	Prism &	1.00 &	hyades64 (1.10), la93-101(1.20), \\
& & & & & & & & la97-249(1.10) \\
135 Hertha & 08 Nov. 08	& 10:51	& 400 &	IRTF &	SPEX &	Prism &	1.00 &	hyades64 (1.10), la97-249(1.10) \\
135 Hertha & 22 Dec. 08	& 06:35	& 870 &	IRTF &	SPEX &	Prism &	1.05 & hyades64(1.05), la93-101(1.10) \\
161 Athor     &12 Aug 05 & 04:29 & 160 & NTT & SOFI & GBF  &1.06  & hip096165  (1.09) \\
161 Athor     &12 Aug 05 & 04:34 & 160 & NTT & SOFI & GRF  &1.07  & hip096165 (1.09) \\      	
161 Athor     &13 Aug 05&02:47&   150& NTT & EMMI &	GR1 &	1.01 & HD144585 (1.22) \\
201 Penelope  & 29 Feb. 04 & 00:32 & 90 & TNG & Dolores & LR-R & 1.15 & hd89010 (1.02)   \\
201 Penelope  & 29 Feb. 04 & 00:35 & 90 & TNG & Dolores & LR-B & 1.16 & hd89010 (1.18)   \\
216 Kleopatra & 18 Nov. 04 & 05:25 & 90 &  TNG & Dolores & LR-R & 1.09 & la98-978 (1.20) \\ 
216 Kleopatra & 18 Nov. 04 & 05:28 & 90 &  TNG & Dolores & MR-B & 1.09 & la98-978 (1.20) \\ 
216 Kleopatra & 19 Nov. 04 & 04:45 & 180 & TNG & NICS & Amici   & 1.10  & la98-978 (1.17) \\
224 Oceana & 15 Nov. 04 & 21:24 & 240 & TNG & Dolores & LR-R & 1.08 & hyades64 (1.03) \\
224 Oceana & 15 Nov. 04 & 21:29 & 240 & TNG & Dolores & MR-B & 1.08 & hyades64 (1.03) \\
224 Oceana & 20 Sep. 05 &	15:20 &	960  &	IRTF &	SPEX &	Prism &	1.20 & la110-361(1.10), la115-270(1.10), \\
& & & & & & & & la93-101(1.1), la112-1333(1.1) \\
250 Bettina &18 Nov. 04 & 06:46 & 120 &  TNG & Dolores & LR-R & 1.04 & hyades64 (1.40) \\ 
250 Bettina & 18 Nov. 04 & 06:49 & 120 &  TNG & Dolores & MR-B & 1.05 & hyades64 (1.40) \\ 
250 Bettina & 21 Nov. 04 & 06:41 & 80 & TNG & NICS & Amici   & 1.04  & la98-978 (1.17)  \\
325 Heidelberga & 20 Nov. 04 & 01:42 & 60 & TNG & Dolores & LR-R  & 1.15 &  hyades64 (1.02)  \\
325 Heidelberga & 20 Nov. 04 & 01:45 & 120 & TNG & Dolores & MR-B  & 1.16 &  hyades64 (1.02)  \\
325 Heidelberga & 19 Nov. 04 & 00:24 & 600 & TNG & NICS & Amici   & 1.02  & hyades64 (1.03) \\
338 Budrosa   &13 Aug 05&01:17	&360& NTT & EMMI &	GR1 &	1.10  & HD144585 (1.05) \\       	
338 Budrosa   &12 Aug 05 & 01:40 & 360 & NTT & SOFI & GBF  &1.14  & hip083805  (1.23) \\ 
338 Budrosa   &12 Aug 05 & 02:10 & 280 & NTT & SOFI & GRF  &1.22  & hip083805  (1.23) \\ 
347 Pariana   &13 Aug 05 &06:57	&420& NTT & EMMI &	GR1 &	1.02  & HD1835 (1.07) \\         	
347 Pariana   &12 Aug 05 &07:30 &480 & NTT & SOFI & GBF  &1.04  &  hip103579  (1.06) \\
347 Pariana   &12 Aug 05 &07:39 &480 & NTT & SOFI & GRF  &1.05  &  hip103579  (1.06) \\
369 Aeria     &13 Aug 05 &01:08	&240& NTT & EMMI &	GR1 &	1.07 & HD144585 (1.05) \\          	
369 Aeria     &12 Aug 05 & 02:49&240& NTT & SOFI & GBF  &1.33  & hip083805  (1.23) \\ 
369 Aeria     &12 Aug 05 & 02:56&280& NTT & SOFI & GBF  &1.36  & hip083805  (1.23) \\ 
382 Dodona    &13 Aug 05 & 00:34&300& NTT & EMMI &	GR1 &	1.45 & HD144585 (1.22) \\    	 	
418 Alemania  & 20 Jan 07 & 09:05 & 360 & NTT & EMMI & GR1 & 1.13 &  la98-978 (1.22) \\
441 Bathilde   &13 Aug 05 &05:58	&180& NTT & EMMI &	GR1 &	1.21  & HD1835 (1.07) \\         
441 Bathilde   &12 Aug 05 &04:54 &280& NTT & SOFI &      GBF &   1.10  & hip096165  (1.09)  \\
441 Bathilde   &12 Aug 05 &05:01 &280& NTT & SOFI &      GRF &   1.10  & hip096165  (1.09)  \\
498 Tokio     &13 Aug 05 &05:45	& 60 & NTT & EMMI &	GR1 &	1.10 & HD1835 (1.07) \\
498 Tokio     &12 Aug 05 &05:37 &120 & NTT & SOFI &     GBF &   1.07  & hip103572 (1.06) \\
498 Tokio     &12 Aug 05 &05:40 &160 & NTT & SOFI &     GBF &   1.08  & hip103572 (1.06) \\
516 Amherstia &  19 Nov. 04 & 00:07 & 360 & TNG & NICS & Amici   & 1.20  & la98-978 (1.17) \\
516 Amherstia & 20 Jan 07 & 09:23 & 60 & NTT & EMMI & GR1 & 1.07 &  la102-1081 (1.22) \\
558 Carmen    &13 Aug 05 &00:50	&480& NTT & EMMI &	GR1 &	1.49  & HD144585 (1.22) \\       	
755 Quintilla &13 Aug 05 &02:32	&600& NTT & EMMI &	GR1 &	1.08  & HD144585 (1.07) \\      	
755 Quintilla &14 Aug 05 &02:56 &600& NTT & SOFI &      GBF &   1.03  & hip99046 (1.01) \\
755 Quintilla &14 Aug 05 &03:19 &960& NTT & SOFI &      GRF &   1.05  & hip99046 (1.01) \\
785 Zwetana  & 29 Feb. 04 & 22:06 & 240 & TNG & Dolores & LR-R & 1.46 & la102-1081 (1.34)   \\
785 Zwetana  & 1 Mar. 04 & 22:02 & 960 & TNG & NICS & Amici   & 1.39  & hyades64 (1.21) \\
785 Zwetana   &12 Aug 05 &23:03 &240& NTT & EMMI &	GR1 &	1.08  & HD144585 (1.05) \\       	
849 Ara & 15 Nov. 04 & 20:10 & 120 & TNG & Dolores & LR-R & 1.09 & HD28099 (1.05) \\
849 Ara & 15 Nov. 04 & 20:19 & 120 & TNG & Dolores & LR-B & 1.10 & hyades64 (1.03) \\
849 Ara       &12 Aug 05 &09:55 &480 & NTT & SOFI &     GBF &   1.73  & hip11355 (1.49) \\
849 Ara       &12 Aug 05 &10:05 &480 & NTT & SOFI &     GRF &   1.69  & hip11355 (1.49) \\ 
860 Ursina  & 20 Nov. 04 & 02:03 & 180 & TNG & Dolores & LR-R  & 1.15 &  hyades64 (1.02)  \\
860 Ursina  & 20 Nov. 04 & 02:07 & 240 & TNG & Dolores & MR-B  & 1.16 &  hyades64 (1.02)  \\
860 Ursina  & 18 Nov. 04 & 22:48 & 480 & TNG & NICS & Amici   & 1.03  & hyades64 (1.03) \\
872 Holda & 20 Nov. 04 & 21:34 & 300 & TNG & Dolores & LR-R  & 1.18 & la115-271 (1.16)      \\
872 Holda & 20 Nov. 04 & 21:40 & 300 & TNG & Dolores & MR-B  & 1.19 & la115-271 (1.16)   \\
872 Holda & 16 Nov. 05 & 12:51 & 600 & IRTF & SPEX & Prism &1.01 & hyades64 (1.15), \\
& & & & & & & & la115-270(1.10), la93-101(1.20)\\
\hline
\end{longtable}
\end{center}
}
\begin{list}{}{}
\item 
\end{list}


       \begin{sidewaystable}
       \caption{Physical and orbital parameters of the M asteroids
observed and selected on the basis of the Tholen taxonomy (Tholen, 1984). The
Bus and Bus-DeMeo classifications are 
also reported, together with the spectral slopes value (S$_{VIS}$ calculated 
for the 0.55-0.8
$\mu$m wavelength range, S$_{NIR1}$ for the 1.1-1.6 $\mu$m range, S$_{NIR2}$
for the 1.7-2.4 $\mu$m range, and S$_{cont}$ for the whole range).}
        \label{slope}
        \scriptsize{
\begin{tabular}{|l|c|c|c|c|c|c|c|c|c|c|c|c|} \hline
Asteroid & Bus &  Bus-Demeo & albedo &  D  & Rot  & a
 & e &    i & S$_{Vis}$   & S$_{NIR1}$ & S$_{NIR2}$ & S$_{cont}$  \\ 
 &  &  &  &  (Km) & (h)  & (UA) &  &    ($^{o}$) & (\%/$10^{3}$\AA)   &  (\%/$10^{3}$\AA)
& $(\%/10^{3}$\AA) & $(\%/10^{3}$\AA) \\ \hline
16 Psyche        & X  &Xk   &0.12  &253.16  &4.196 & 2.9201 & 0.1395 &3.1  &   
3.00$\pm$0.54&  3.79$\pm$0.93&  4.26$\pm$0.84 &  3.66$\pm$0.71 \\
22 Kalliope	 & X  &Xk   &0.14  &181.00  &4.148 & 2.9077 & 0.1030 &13.7 &   
3.85$\pm$0.55&  2.59$\pm$0.75&  1.61$\pm$0.72 &  2.88$\pm$0.72 \\
55 Pandora	 & X  &Xk   &0.30  &66.70   &4.804 & 2.7582 & 0.1447 &7.2  &   
3.19$\pm$0.55&  2.25$\pm$0.75&  1.90$\pm$0.73 &  2.25$\pm$0.71 \\
69 Hesperia	 & X  &Xk   &0.14  &138.13  &5.656 & 2.9776 & 0.1686 &8.6  &   
4.06$\pm$0.55&  2.81$\pm$0.81& -1.33$\pm$0.91 &  3.09$\pm$0.72 \\
97 Klotho	 & $-$ &Xc  &0.23  &82.83   &35.15 & 2.6674 & 0.2571 &11.8 &   
3.19$\pm$0.53&  1.07$\pm$0.73& -0.36$\pm$0.72 &  1.23$\pm$0.72 \\
110 Lydia	 & X  &Xk   &0.18  &86.09   &10.92 & 2.7321 & 0.0799 &5.9  &   
3.94$\pm$0.54&  2.75$\pm$0.87&  1.21$\pm$0.89 &  2.88$\pm$0.72 \\
125 Liberatrix 	 & X  &$-$  &0.22  &43.58   &3.968 & 2.7433 & 0.0807 &4.6  &   
1.60$\pm$0.55&  --&  -- &  1.63$\pm$0.73 \\
129 Antigone	 & X  &Xc   &0.16   &113.00  &4.957 & 2.8674 & 0.2136 &12.2 &   
3.49$\pm$0.57&  0.56$\pm$0.74&  1.06$\pm$0.77 &  0.92$\pm$0.71 \\
132 Aethra	 & Xe &$-$  &0.20  &42.87   &5.168 & 2.6082 & 0.3887 &25.1 &   
2.35$\pm$0.59&  --&  -- &  2.83$\pm$0.78 \\
135 Hertha	 & Xk &Xk   &0.14  &79.24   &8.400 & 2.4286 & 0.2063 &2.3  &   
3.22$\pm$0.55&  1.87$\pm$0.73&  1.25$\pm$0.71 &  2.07$\pm$0.72 \\
161 Athor	 & Xc &Xc   &0.20  &44.19   &7.288 & 2.3805 & 0.1374 &9.0  &   
2.18$\pm$0.66&  1.39$\pm$0.74&  0.14$\pm$0.73 &  1.16$\pm$0.71 \\
201 Penelope	 & X  &$-$  &0.16  &68.39   &3.747 & 2.6790 & 0.1809 &5.7  &   
5.57$\pm$0.57&  --&  -- &  5.53$\pm$0.74 \\
216 Kleopatra	 & Xe &Xk   &0.12  &124.00  &5.385 & 2.7933 & 0.2521 &13.1 &   
5.40$\pm$0.53&  2.96$\pm$0.80&  1.94$\pm$0.86 &  3.36$\pm$0.73 \\
224 Oceana	 & $-$ &Xc  &0.17  &61.82   &9.385 & 2.6454 & 0.0455 &5.8  &   
1.47$\pm$0.56&  1.69$\pm$0.79&  0.51$\pm$0.74 &  2.18$\pm$0.71 \\
250 Bettina	 & Xk &Xk   &0.26  &79.75   &5.054 & 3.1497 & 0.1245 &12.8 &   
3.79$\pm$0.56&  7.00$\pm$0.91&  4.22$\pm$0.82 &  4.83$\pm$0.72 \\
325 Heidelberga	 & $-$ &X   &0.11  &75.72   &6.737 & 3.204  & 0.1686 &8.5  &   
5.67$\pm$0.54&  3.82$\pm$0.83&  2.10$\pm$0.79 &  4.26$\pm$0.72 \\
338 Budrosa	 & Xk &Xk   &0.18  &63.11   &4.6	& 2.9124 & 0.0208 &6.0  
&    4.15$\pm$0.60&  3.11$\pm$0.79&  3.29$\pm$0.78 &  3.12$\pm$0.71 \\
347 Pariana	 & $-$ &Xk  &0.18  &51.36   &4.052 & 2.6153 & 0.1639 &11.7 &   
3.66$\pm$0.55&  1.86$\pm$0.76&  1.97$\pm$0.75 &  2.32$\pm$0.71 \\
369 Aeria	 & $-$ &Xk  &0.19  &60.00   &4.787 & 2.6480 & 0.0985 &12.7 &   
4.48$\pm$0.57&  3.18$\pm$0.75&  1.92$\pm$0.77 &  2.75$\pm$0.71 \\
382 Dodona	 & $-$ &$-$ &0.16  &58.37   &4.116 & 3.1183 & 0.1761 &7.4  &   
5.02$\pm$0.58&  --&  -- &  4.76$\pm$0.75 \\
418 Alemania	 & X  &$-$  &0.19  &34.10   &4.671 & 2.5919 & 0.1195 &6.8  &   
4.07$\pm$0.55&  --&  -- &  3.39$\pm$0.73 \\
441 Bathilde	 & Xk &X    &0.14  &70.32   &10.44 & 2.8049 & 0.0823 &8.1  &   
5.30$\pm$0.57&  2.21$\pm$0.74&  0.20$\pm$0.74 &  2.12$\pm$0.72 \\
498 Tokio	 & $-$ &Xc  &0.07  &81.83   &20.00 & 2.6513 & 0.2245 &9.5  &   
5.76$\pm$0.57&  0.06$\pm$0.75&  0.20$\pm$0.73 &  0.49$\pm$0.72 \\
516 Amherstia	 & X  &Sq   &0.16  &73.10   &7.49  & 2.6788 & 0.2737 &12.9 &   
3.46$\pm$0.55&  1.95$\pm$0.86&  1.04$\pm$0.87 &  1.51$\pm$0.73 \\
558 Carmen	 & Xk &$-$  &0.12  &59.31   &11.38 & 2.9080 & 0.0433 &8.4  &   
3.47$\pm$0.60&  --&  -- &  3.34$\pm$0.78 \\
755 Quintilla	 & Xk &Xk   &0.16  &36.04   &4.551 & 3.1729 & 0.1460 &3.3  &   
3.82$\pm$0.59&  2.70$\pm$0.77&  0.38$\pm$0.77 &  2.66$\pm$0.73 \\
785 Zwetana	 & Cb &X    &0.12  &48.54   &8.888 & 2.5695 & 0.2099 &12.7 &   
4.03$\pm$0.57&  2.60$\pm$0.81&  1.80$\pm$0.91 &  2.78$\pm$0.72 \\
849 Ara		 & $-$ &D  &0.27  &61.82   &4.116 & 3.1526 & 0.1961 &19.5  &   
4.50$\pm$0.57&  7.96$\pm$0.82&  3.78$\pm$0.82 &  5.48$\pm$0.72 \\
860 Ursina	 & X  &X    &0.16  &29.32   &9.386 & 2.7947 & 0.1078 &13.3 &   
4.00$\pm$0.57&  3.98$\pm$0.82&  3.41$\pm$0.80 &  3.43$\pm$0.71 \\   
872 Holda	 & X  &Xk  &0.21  &30.04   &7.20  & 2.7305 & 0.0801 &7.3   &   
1.29$\pm$0.56&  2.46$\pm$0.76&  1.35$\pm$0.72 &  1.94$\pm$0.71 \\ \hline
\end{tabular}
}
\end{sidewaystable}


       \begin{table}
       \begin{center}
       \caption{Band center, depth and width for the features detected on the
asteroid spectra. The results of the 3 $\mu$m band investigation from Rivkin et
al. (1995, 2000) are also reported.}
        \label{band}
\begin{tabular}{|l|c|c|c|c|} \hline
Asteroid & Band center ($\mu$m) & Depth (\%)  & Width ($\mu$m) & 3 $\mu$m band
\\ \hline
16 Psyche         &0.949$\pm$0.008  &2.9  &0.898--1.020 & No  \\
16 Psyche         &0.430$\pm$0.004  &1.6  &0.418--0.442 & No  \\
22 Kalliope	  &0.903$\pm$0.008  &2.5  &0.756--1.017 & Yes  \\
22 Kalliope	  &0.434$\pm$0.005  &1.2  &0.407--0.456 & Yes  \\
55 Pandora	  &0.910$\pm$0.010  &1.7  &0.790--0.970 & Yes  \\
69 Hesperia	  &0.951$\pm$0.009  &2.6  &0.830--1.080 & --  \\
69 Hesperia	  &0.430$\pm$0.004  &1.5  &0.417--0.443 & --  \\
110 Lydia	  &0.942$\pm$0.008  &2.0  &0.857--1.033 & Yes \\
125 Liberatrix    &  --             & --  & --          & No \\
129 Antigone      &1.028$\pm$0.010  & 1.5 &0.929--1.136 & Yes \\
132 Aethra	  &0.498$\pm$0.004  &2.0  &0.456--0.537 & --  \\
135 Hertha	  &0.905$\pm$0.008  &1.9  &0.771--1.042 & Yes \\
135 Hertha	  &0.515$\pm$0.005  &1.2  &0.497--0.531 & Yes \\
161 Athor         &  --             & --  & --          & No \\
201 Penelope      &  --             & --  & --          & Yes \\
216 Kleopatra	  &0.969$\pm$0.008  &3.2  &0.815--1.104 & No  \\
216 Kleopatra	  &0.429$\pm$0.004  &3.2  &0.411--0.446 & No \\
250 Bettina	  &0.885$\pm$0.010  &2.0  &0.819--0.972 & -- \\
338 Budrosa	  &0.876$\pm$0.010  &2.1  &0.766--0.981 & -- \\
338 Budrosa	  &0.425$\pm$0.004  &4.2  &0.407--0.446 & --  \\
347 Pariana	  &0.871$\pm$0.008  &2.4  &0.780--0.950 & -- \\
369 Aeria	  &0.884$\pm$0.008  &1.9  &0.772--0.987 & No? \\
498 Tokio	  &0.430$\pm$0.005  &1.6  &0.413--0.449 & -- \\
498 Tokio	  &1.159$\pm$0.008  &2.5  &0.905--1.384 & -- \\
516 Amherstia	  &0.965$\pm$0.008  &6.8  &0.750--1.440 & -- \\
516 Amherstia	  &1.949$\pm$0.010  &6.1  &1.702--2.193 & -- \\
755 Quintilla	  &0.904$\pm$0.010  &4.9  &0.760--1.010 & -- \\
755 Quintilla	  &1.864$\pm$0.010  &3.7  &1.699--2.062 & -- \\
755 Quintilia	  &1.610$\pm$0.008  &1.9  &1.535--1.717 & --  \\
755 Quintilia     &1.369$\pm$0.010  &3.0  &1.237--1.507 & -- \\
785 Zwetana       &  --             & --  & --          & No \\
872 Holda	  &0.965$\pm$0.020  &2.8  &0.866--1.212 & -- \\
 \hline
\end{tabular}
\end{center}
\end{table}

\newpage
       \begin{table}[h]
       \begin{center}
       \small{
       \caption{Summary of the features observed on the asteroids spectra and
possible mineralogical interpretation.The Bus-DeMeo (BD) taxonomy is also reported. CFT=crystal field transitions.}
        \label{features}
\begin{tabular}{|l|c|l|l|} \hline
band   & asteroids & BD  & suggested mineralogical interpretation \\
($\mu$m) & & tax. & \\ \hline
0.43          & 16, 22, 69, 216, 338 & Xk  &  chlorites and Mg-rich
serpentines\\
                       &     & & or pigeonite or augite \\
0.43          & 498 & Xc &      iron sulfate jarosite \\

0.49          &  132   & - & sulfides such as oldhamite and/or troilite \\
0.51          & 135    & Xk & Fe$^{2+}$ spin-forbidden CFT in pyroxene \\
0.90      & 16, 22, 55, 69, 110, 135, 216, &  Xk       & low-Fe, low-Ca
orthopyroxene minerals \\
           & 250, 338, 347, 369, 755, 872  &  Xk &  \\
0.93-1.14    & 129 & Xc   & antigorite? \\
0.9-1.3    & 498 & Xc   & unknown \\
1.37       & 755 & Xk & unknown \\
1.61       & 755 & Xk & unknown \\
1.86       & 755 & Xk & iron bearing pyroxene \\ 
0.96 and 2 & 516 & Sq & olivine and pyroxene \\
\hline
\end{tabular}
}
\end{center}
\end{table}

\hspace{10cm}

\begin{sidewaystable}
       \caption{The best matches between the 24 asteroid sample observed in the
visible and near infrared range and meteorites from the RELAB database.  $\chi^{2}$ values noted with an asterisk indicate asteroids for which the best RELAB fit matches only the slope and albedo but not the 0.9 $\mu$m feature.}
        \label{tab}
        \scriptsize{
\begin{tabular}{|l|c|c|c|c|c|l|l|} \hline
Asteroid & Albedo & RELAB file & Met. name & Met. Class & Met. Albedo & Grain Size ($\mu$m) &
$\chi^{2}$ value \\ \hline
16 Psyche             & 0.12 & eac/sc/s1sc99 & MET101A & IM	     & 0.13 &
particulate & 6.46667e-04$^{*}$ \\
22 Kalliope           & 0.14 & txh/mb/c1mb46 & Landes & IM            & 0.16 & slab      
 & 1.84672e-03 \\
55 Pandora            & 0.30 & txh/mb/cmmb43 & Esquel & Pall         & 0.29 & sanded slab
& 4.15916e-03 \\
69 Hesperia           & 0.14 & eac/sc/lasc99 & MET101A & IM           & 0.12 & particulate & 8.74344e-04$^{*}$ \\
97 Klotho             & 0.23 & mjg/mr/cgp017 & Babb's Mill & IM          & 0.24 & $<$300  &
5.14856e-04 \\
110 Lydia             & 0.18 & txh/mb/ckmb43 & Esquel & Pall        & 0.14 & $<$63   &
4.52068e-04$^{*}$ \\
129 Antigone          & 0.16 & eac/sc/c1sc14 & AWA101 & Iron-Nickel & 0.11 & 45-90 &
3.05211e-03 \\
135 Hertha            & 0.14 & txh/mb/c1mb46 & Landes & IM          & 0.17 & slab  &
8.47345e-04 \\
161 Athor             & 0.20 & mjg/mr/mgn021 & Pillistfer & EC	     & 0.17 & $<$300  &
1.46184e-03 \\
216 Kleopatra         & 0.12 & txh/mb/cfmb47 & DRP78007 & IM           & 0.11 & $<$25      
  & 1.70951e-03 \\
224 Oceana            & 0.17 & txh/mb/c1mb46 & Landes & IM          & 0.16 & slab  &
1.26472e-03 \\
250 Bettina           & 0.26 & txh/mb/r4mb39 & Y-8451,20 & Pall	     & 0.25 & --     &
1.18340e-03$^{*}$ \\
325 Heidelberga       & 0.11 & cmp/mi/c1mi08 & Brown (filings) & IM          & 0.15 & filings &
7.68864e-04 \\
338 Budrosa           & 0.18 & eac/sc/lasc99 & MET101A  & IM	     & 0.13 &
particulate & 1.31101e-03\\
347 Pariana           & 0.18 & txh/mb/c1mb46 & Landes & IM	     & 0.17 & slab  &
9.04683e-04 \\
369 Aeria             & 0.19 & eac/sc/c1sc99 & MET101A & IM          & 0.13 & particulate
& 1.98982e-03$^{*}$\\
441 Bathilde          & 0.14 & atb/ma/c1ma47 & -- & Laser Irr Ol & 0.12 & $>$125 &
2.18014e-03 \\
498 Tokio             & 0.07 & txh/mb/ncmb59 & Vigarano & CV	     & 0.08 & $<$100  &
1.27790e-03 \\
516 Amherstia         & 0.16 & atb/ma/c1ma53 & Tsarev & OC*	     & 0.14 & $>$300  &
5.49437e-04 \\
755 Quintilla         & 0.16 & txh/mb/cfmb47 & DRP78007 & IM            & 0.12 & $<$25 & 
1.15605e-02$^{*}$\\
785 Zwetana           & 0.12 & txh/mb/ckmb43 & Esquel & Pall        & 0.14 & $<$63   &
8.76718e-04 \\
849 Ara               & 0.27 & tjm/tb/c1tb56 & Metal in ALH84019 & Iron-Nickel  & 0.22 & thin
section &4.08895e-03 \\
860 Ursina            & 0.17 & eac/sc/c1sc99 & MET101A & IM	     & 0.13 &
particulate & 6.02053e-04 \\ 
872 Holda             & 0.22 & txh/mb/c1mb46 & Landes &  IM	     & 0.16 & slab   & 2.03065e-03$^{*}$ \\ \hline 
\end{tabular}
}
\end{sidewaystable}


\begin{table} 
\caption{Significant Spearman Correlations (the most significant ones are in
bold).}
\label{stat}
\begin{tabular}{|l|c|c|c|} \hline
  & $\rho$ & $P_{r}$ & $n$\\ \hline
{\boldmath $S_{cont} ~vs~ a ~ (all)$}                        &  {\bf
0.638498}  & {\bf 0.00104248}   & {\bf 23} \\
$S_{cont}$ vs $a$ (with the 0.9$\mu$m band)     & 0.646493   &  0.0169538   &
13 \\
$S_{cont}$ vs $a$ (without the 0.9$\mu$m band)  &  0.430303  &   0.214492 & 10
\\\hline
$S_{VIS}$ vs $a$ (all)                        &  0.311615  &  0.0998569   & 29
\\
$S_{VIS}$ vs $a$ (with the 0.9$\mu$m band)     &  0.148352 &    0.628610    &
13 \\
$S_{VIS}$ vs $a$ (without the 0.9$\mu$m band)  &   0.426471  &  0.0994991 & 16
\\ \hline
{\boldmath $S_{NIR1} ~ vs~ a ~(all)$}        &  {\bf 0.630435} &  {\bf
0.00126108}  & {\bf 23} \\
$S_{NIR1}$ vs $a$ (with the 0.9$\mu$m band)     &  0.571429 &   0.0413423  &
13 \\
$S_{NIR1}$ vs $a$ (without the 0.9$\mu$m band)  &  0.642857 &   0.0855589  &
10 \\ \hline
$S_{NIR2}$ vs $a$ (all)                        &  0.331604 &   0.122167   & 23
\\
$S_{NIR2}$ vs $a$ (with the 0.9$\mu$m band)     &  0.0494505 &    0.872543 &
13 \\
$S_{NIR2}$ vs $a$ (without the 0.9$\mu$m band)  & 0.498483 &   0.142518   & 10
\\ \hline
{\boldmath $S_{cont} ~ vs ~rot ~ (all)$}          & {\bf -0.418088}  & {\bf
0.0471146}   & {\bf 23} \\
$S_{cont}$ vs $rot$ (with the 0.9$\mu$m band)          &   -0.162311  &  
0.596258 & 13 \\
$S_{cont}$ vs $rot$ (without the 0.9$\mu$m band)       &  -0.406061  &  
0.244282  & 10 \\ \hline
{\boldmath $S_{NIR1} ~ vs ~rot ~(all)$}           &   {\bf-0.482213} & 
{\bf0.0197941}     & {\bf23} \\
$S_{NIR1}$ vs $rot$ (with the 0.9$\mu$m band)   &  -0.0274725 &    0.929013 &
13 \\
$S_{NIR1}$ vs $rot$ (without the 0.9$\mu$m band) &  -0.466667  &   0.173939  &
10 \\ \hline
{\boldmath $S_{NIR2} ~ vs ~rot ~ (all)$}        & {\bf-0.516926}  &  {\bf
0.0115455}       & {\bf23} \\
$S_{NIR2}$ vs $rot$ (with the 0.9$\mu$m band)   &  -0.478022 &   0.0984901  &
13 \\
$S_{NIR2}$ vs $rot$ (without the 0.9$\mu$m band)& -0.547115  &   0.101678  &
10 \\ \hline
{\boldmath $S_{VIS} ~ vs ~ alb ~ (all)$}     & {\bf -0.486513}  & {\bf
0.00744684}  & {\bf 29} \\
$S_{VIS}$ vs $alb$ (with the 0.9$\mu$m band)   & -0.329670  &   0.271335 & 13
\\
$S_{VIS}$ vs $alb$ (without the 0.9$\mu$m band)   & -0.635294  &  0.00818205 &
16 \\
\hline
\end{tabular}
\end{table}

\hspace{3cm}
\newpage

{\bf Figure captions}

\vspace{1cm}

Fig. 1 - Visible spectra of 6 M-type asteroids. For 125 Liberatrix three spectra
were acquired on 2004, November 20: the spectrum labelled A corresponds to
UT=20:37, the B to UT=22:01, and the C to UT=23:27.

Fig. 2 - Observed spectra for Tholen M-type asteroids extended into the near-infrared.

Fig. 3 - Observed spectra for Tholen M-type asteroids extended into the near-infrared.

Fig. 4 - Observed spectra for Tholen M-type asteroids extended into the near-infrared.

Fig. 5 - Observed spectra for Tholen M-type asteroids extended into the near-infrared.

Fig. 6 - Spectral matches between the featureless observed asteroids (in black) and
meteorites (in red) from the RELAB database (see 
Table~\ref{tab} for details on the meteorites samples).

Fig. 7 - Spectral matches between the observed asteroids (in black) showing the 0.43 $\mu$m feature and
meteorites (in red) from the RELAB database (see 
Table~\ref{tab} for details on the meteorites samples). In the Bus-Demeo taxonomy all these asteroids, except 498 Tokio (Xc-type), belong to the Xk-type and show also a band in the 0.9 $\mu$m region  

Fig. 8 - Spectral matches between the observed asteroids (in black) having absorption features beyond 0.8 $\mu$m and meteorites (in red) from the RELAB database (see 
Table~\ref{tab} for details on the meteorites samples). All asteroids belong to the Xk-type in the Bus-Demeo taxonomy, except 129 Antigone (Xc-type), and 516 Amherstia (Sq-type).

Fig. 9 - Visual and NIR1 slopes versus the semi-major axis, 
albedo and rotational period for the observed asteroids. Black bullet
represent the asteroids showing a band in the 0.9 $\mu$m region and belonging to
the Xk-type in the Bus-DeMeo taxonomy.

Fig. 10 - Geographycal mixtures proposed to interpret the surface composition of
22 
Kalliope: 98\% pallasite and 2\% goethite (continuous line, mixture albedo =
0.14); 99\% pallasite and 1.0\% orthopyroxene (dashed line, mixture 
albedo = 0.14).

{\bf Figures}

\end{document}